# Excitation frequency dependence of noise and minimum detectable force in amplitude-modulation atomic force microscopy

Kenichi Umeda[1,2] and Noriyuki Kodera[1]

[1] *Nano Life Science Institute (WPI-NanoLSI),*

*Kanazawa University, Kakuma-machi, Kanazawa, Ishikawa, 920-1192, Japan.*

[2] *PRESTO/JST, 4-1-8 Honcho, Kawaguchi, Saitama 332-0012, Japan.*

Corresponding Authors

Dr. Kenichi Umeda (E-mail: umeda.k@staff.kanazawa-u.ac.jp)

Prof. Noriyuki Kodera (E-mail: nkodera@staff.kanazawa-u.ac.jp)

## Abstract

Atomic force microscopy (AFM) is a versatile nanoscale imaging technique. Since its spatiotemporal resolution is fundamentally limited by the minimum detectable force (MDF) arising from system noise, a deep understanding of MDF is essential for improving instrumentation. However, the theoretical MDF of amplitude-modulation (AM) AFM has long remained inconsistent, with three reported expressions yielding conflicting coefficients: 1.84, 1.41, and 0.71 times those of other dynamic modes. Moreover, although we recently clarified the strong dependence of force sensitivity on the cantilever's driving frequency, previous theories overlooked this effect. Here, we present an exact solution for the MDF of AM-AFM that accounts for noise frequency dependence, excitation efficiency, and arbitrary cantilever Q-factors. Our results clarify that the coefficient strongly depends on the driving frequency and Q-factor. Notably, when driven at the resonance slope, it stays within 1 and 1.41, thus resolving this long-standing inconsistency. Our findings provide essential guidance for improving instrumentation to visualize previously inaccessible phenomena.

## 1. Introduction

Atomic force microscopy (AFM) is a powerful technique for visualizing the nanoscale surface topography [1-9]. Owing to its capability to perform operando measurements with spatial resolution beyond the reach of other techniques under a variety of environments—including vacuum, air, and liquid—it has been applied across a wide range of disciplines [3,9-17]. Beyond topographical imaging, AFM uniquely detects forces acting on the probe, which not only serve as feedback signals but also enable its use as a sensitive force sensor or mechanical stimulator to investigate the mechanistic and viscoelastic properties of materials [9,18-20] and supramolecular mechanisms [17,21-24]. Accordingly, accurate force quantification is of central importance.

However, the accuracy and resolution of force measurements are fundamentally limited by the minimum detectable force (MDF), defined as the lowest force detectable at a signal-to-noise ratio (SNR) of 1. Since MDF is closely linked to spatiotemporal resolution, improving it has unlocked access to previously inaccessible phenomena, e.g., single-molecule chemical structure imaging in vacuum [16], atomic-scale three-dimensional hydration imaging in liquids [13,25-27], and video-rate submolecular-scale imaging of biomolecules in action [15,28,29].

MDF is also a widely recognized performance metric for various types of AFM instrumentation [1,3,5-8,28,30-45] and other force-sensing techniques [34,46-49]. As MDF varies significantly depending on cantilever geometry and imaging conditions, it serves as a critical guide for improving instrumentation [8,35,36,45,50-54]. Therefore, a deep understanding of MDF is essential for driving new technological innovations in AFM.

While AFM techniques are classified by their force detection mechanisms—amplitude-modulation (AM) [1-3,6,55-57], frequency-modulation (FM) [6,7,16,25,27,30,39,58], and phase-modulation (PM) [7,31-33] modes—equivalent theoretical expressions for MDF have been reported for FM- and PM-AFM. In contrast, for AM-AFM, three theoretical expressions have been reported that share the same form as those for the other modes but have distinct coefficients: 1.84 (= $\sqrt{27/4}$ ), 1.41 (= $\sqrt{2}$ ), or 0.71 (= $\sqrt{1/2}$ ) times as large [1,2,5,30,57]. These discrepancies raise questions as to why AM-AFM alone yields a different MDF compared to the other modes, whether it should be worse or better, and whether these expressions are indeed valid.

Furthermore, we previously demonstrated that force detection sensitivity in AM-AFM strongly depends on the cantilever's driving frequency ($f_{\mathrm{drive}}$), and unlike in the other modes, reaches its maximum when excited at the resonance slope rather than at the resonance frequency ($f_0$) [59]. Nonetheless, excitation at $f_0$ remains common due to its simplicity and theoretical convenience [12,60-62]. Off-resonance excitation combined with peak force detection has also recently been proposed to minimize dynamic impact forces [17]. Despite the variety of excitation strategies in AM-AFM, no prior study has comprehensively evaluated the optimal $f_{\mathrm{drive}}$ from the perspective of minimizing MDF.

In previous studies, force has been formulated based on the simple harmonic oscillator (SHO) model [4,12,13,59,63]; however, this model is known to break down under highly viscous environments [64-69]. The systematic error caused by this model mismatch in force detection has not been investigated, leaving uncertainty regarding the accuracy of force measurements.

In this study, we present a comprehensive analysis of AM noise and MDF in AM-AFM, focusing on their dependence on the $f_{\mathrm{drive}}$ and cantilever Q-factor ($Q_{\mathrm{cl}}$) through simulations and experiments. We show that the MDF decreases as $f_{\mathrm{drive}}$ is lowered, approaching the theoretical dynamic-mode limit imposed by cantilever Brownian motion noise. The exact MDF theory developed here offers a general framework for benchmarking system performance.

## 2. Force gradient dependent on driving frequency

As previously studied [59,70], in AM-AFM, the average tip–sample interaction force ($\langle F_{\text{ts}} \rangle$) can be estimated from amplitude reduction based on a large-amplitude approximation. However, in contrast to static-mode AFM, dynamic-mode AFM is more likely to detect the resonance shift resulting from the force gradient ($k_{\text{ts}} \equiv F'_{\text{ts}}$), rather than the force magnitude. Consequently, theoretical equations for the minimum detectable $k_{\text{ts}}$ in a small-amplitude limit are often utilized [34-38,45]. To facilitate comparison with previous theories, we first derive analytical solutions for $k_{\text{ts}}$ in AM-AFM.

Using the Fourier series of the tip–sample interaction ($F_{\text{ts}}$), $G_{\text{cl}}$, the transfer function of the cantilever based on a simple harmonic oscillator (SHO) model is obtained as follows [4,59]:

$$G_{\text{cl}}(\tilde{\omega}_{\text{drive}}) = \frac{1}{k_{\text{cl}}} \frac{1}{\left(1 - \tilde{\omega}_{\text{drive}}^2 - I_{\text{even}}\left(z_{\text{tip}}, A_{\text{cl}}\right)\right) + \mathrm{i}\left(\tilde{\omega}_{\text{drive}} / Q_{\text{cl}} + I_{\text{odd}}\left(z_{\text{tip}}, A_{\text{cl}}\right)\right)}, \tag{1}$$

where $A_{\text{cl}}$ and $z_{\text{tip}}$ denote the oscillation amplitude and displacement of the cantilever, respectively, and $k_{\text{cl}}$ denotes the dynamic spring constant of the cantilever, which is typically about 3% greater than the static spring constant [71]. $\tilde{\omega}_{\text{drive}}$ is a normalized $f_{\text{drive}}$ that is defined as follows:

$$\tilde{\omega}_{\text{drive}} \equiv \frac{\omega_{\text{drive}}}{\omega_0} = \frac{f_{\text{drive}}}{f_0}. \tag{2}$$

Hereafter, we denote $f$ and $\omega$ (= $2\pi f$) as a certain frequency and its angular counterpart, respectively. In Eq. (1), $I_{\text{even}}$ and $I_{\text{odd}}$ denote the conservative and dissipation terms, respectively, as defined in Ref [59]. When $I_{\text{odd}}$ can be neglected, $G_{\text{cl}}$ is expressed by

$$G_{\text{cl}}(\tilde{\omega}_{\text{drive}}) = \frac{1}{k_{\text{cl}}} \frac{1}{\left(1 - \tilde{\omega}_{\text{drive}}^2 - \frac{2}{k_{\text{cl}} A_{\text{cl}}^2} \langle F_{\text{ts}} z_{\text{tip}} \rangle\right) + \mathrm{i}\left(\tilde{\omega}_{\text{drive}} / Q_{\text{cl}}\right)}. \tag{3}$$

Therefore, $A_{\text{cl}}$ can be obtained by

$$A_{cl}(\tilde{\omega}_{drive}) = |G_{cl}(\tilde{\omega}_{drive})| F_{drive}$$
$$= \frac{A_0}{\sqrt{\left(1-\tilde{\omega}_{drive}^2 - \frac{2}{k_{cl}A_{cl}^2}\langle F_{ts} z_{tip}\rangle\right)^2 + \left(\tilde{\omega}_{drive}/Q_{cl}\right)^2}}, \tag{4}$$

where $F_{drive}$ and $A_0$ denote the driving force and $A_{cl}$ at $\tilde{\omega}_{drive} = 0$, respectively. Furthermore, by using the free oscillation amplitude ($A_{free}$), $A_{cl}$ can be expressed as follows:

$$A_{cl}(\tilde{\omega}_{drive}) = A_{free}\sqrt{\frac{\left(1-\tilde{\omega}_{drive}^2\right)^2 + \left(\tilde{\omega}_{drive}/Q_{cl}\right)^2}{\left(1-\tilde{\omega}_{drive}^2 - \frac{2}{k_{cl}A_{cl}^2}\langle F_{ts} z_{tip}\rangle\right)^2 + \left(\tilde{\omega}_{drive}/Q_{cl}\right)^2}}. \tag{5}$$

When $A_{free}$ is sufficiently small for $k_{ts}$ to be considered constant during one period of oscillation, a small-amplitude approximation can be employed as follows:

$$\lim_{A_{free}\to 0} F_{ts} \approx \langle F_{ts}\rangle + k_{ts}A_{cl}\cos\omega_{drive}t. \tag{6}$$

Using this equation, we obtain an approximation as follows [5]:

$$\lim_{A_{free}\to 0}\langle F_{ts} z_{tip}\rangle \approx \frac{k_{ts}A_{cl}^2}{2}. \tag{7}$$

Substituting this equation into Eq. (5) and solving for $k_{ts}$ yields the expression for $k_{ts}$ as follows:

$$\lim_{A_{free}\to 0} k_{ts} = -\frac{k_{cl}}{\tilde{A}_{cl}}\left[-\left(1-\tilde{\omega}_{drive}^2\right)\tilde{A}_{cl} \overset{\text{rep}}{\underset{\text{att}}{\pm}} \sqrt{\left(1-\tilde{\omega}_{drive}^2\right)^2 + \left(\tilde{\omega}_{drive}/Q_{cl}\right)^2\left(1-\tilde{A}_{cl}^2\right)}\right], \tag{8}$$

where "rep" and "att" above and below the ± symbol denote repulsive ($\langle F_{ts}\rangle > 0$) and attractive ($\langle F_{ts}\rangle < 0$) forces, respectively. This convention will be followed throughout the paper. $\tilde{A}_{cl}$ is a normalized $A_{cl}$ that is defined as:

$$\tilde{A}_{cl} \equiv \frac{A_{cl}}{A_{free}} = \frac{A_{free} + \Delta A_{ts}}{A_{free}}, \tag{9}$$

where $\Delta A_{ts}$ is the amplitude change resulting from $F_{ts}$. In the absence of feedback error, $\tilde{A}_{cl}$ can be identified with the setpoint ratio.

First, in Fig. 1(a), we analyze the repulsive regime using Eq. (8). Since we used a high-speed

(HS-) AFM system [15,28,29] operated in liquids for experimental validation (as shown later), typical HS-AFM experimental parameters are adopted: $f_0$ = 1 MHz, $k_{cl}$ = 0.1 N/m, and $Q_{cl}$ = 1.5. The overall results indicate that $k_{ts}$ increases as $\tilde{A}_{cl}$ decreases. Notably, at $f_{drive} = f_0$, the equation can be simplified to

$$\lim_{A_{free}\to 0} k_{ts}(\tilde{\omega}_{drive} = 1) = \overset{\text{rep}}{\underset{\text{att}}{\mp}} \frac{k_{cl}}{Q_{cl}} \frac{1}{\tilde{A}_{cl}} \sqrt{1-\tilde{A}_{cl}^2}. \quad (10)$$

As this equation shows, $k_{ts}$ increases steeply near $\tilde{A}_{cl}$ = 1, exhibiting nonlinear behavior.

In contrast, it was found that $k_{ts}$ becomes linear near $\tilde{A}_{cl}$ = 1 when $f_{drive}$ is lower than the peak frequency $f_{peak}$ (0.88 MHz), which is defined as [59],

$$\tilde{\omega}_{peak} \equiv \frac{\omega_{peak}}{\omega_0} = \frac{f_{peak}}{f_0} = \sqrt{1-\frac{1}{2Q_{cl}^2}}, \quad \text{for } Q_{cl} > \frac{1}{\sqrt{2}}. \quad (11)$$

These features are strikingly similar to $\langle F_{ts} \rangle$ formulated in a previous study [59]. However, for $\tilde{A}_{cl}$ < 0.5, unlike $\langle F_{ts} \rangle$, $k_{ts}$ does not saturate and instead diverges infinitely as $\tilde{A}_{cl}$ decreases. Particularly, as indicated by the orange dashed line, the slope near $\tilde{A}_{cl} = 1$ becomes nearly linear at the lower MaxSlope (LMS) frequency, corresponding to the left-side resonance slope [59].

For the attractive regime, the calculations are performed using typical ambient AM-AFM parameters: $f_0$ = 1 MHz, $k_{cl}$ = 3 N/m, and $Q_{cl}$ = 300. In Fig. 1(b), similar to the repulsive regime, near $\tilde{A}_{cl} = 1$, $k_{ts}$ increases steeply at $f_{drive} = f_0$. However, as indicated by the orange dashed line, the curve near $\tilde{A}_{cl} = 1$ becomes nearly linear at the upper MinForce (UMF) and MaxSlope (UMS) frequencies, corresponding to the right-side resonance slope [59], as defined later. Conversely, as $f_{drive}$ rises beyond these frequencies, the increase in $k_{ts}$ becomes steeper.

Therefore, similar to $\langle F_{ts} \rangle$, we found that, when excited at the resonance slope frequencies, the $k_{ts}$ slope near $\tilde{A}_{cl} = 1$ can be accurately represented by a linear approximation as follows:

$$\frac{\partial k_{\text{ts}}}{\partial \tilde{A}_{\text{cl}}} \overset{\text{rep}}{\underset{\text{att}}{=\pm}} \frac{1}{\tilde{A}_{\text{cl}}^2} \frac{k_{\text{cl}}}{Q_{\text{cl}}} \frac{\tilde{\omega}_{\text{drive}}^2 + Q_{\text{cl}}^2\left(1-\tilde{\omega}_{\text{drive}}^2\right)^2}{\sqrt{\left(1-\tilde{A}_{\text{cl}}^2\right)\tilde{\omega}_{\text{drive}}^2 + Q_{\text{cl}}^2\left(1-\tilde{\omega}_{\text{drive}}^2\right)^2}}. \tag{12}$$

By taking the limit of $\tilde{A}_{\text{cl}} \to 1$, we obtain

$$\lim_{\tilde{A}_{\text{cl}}\to 1} \frac{\partial k_{\text{ts}}}{\partial \tilde{A}_{\text{cl}}} = \frac{k_{\text{cl}}}{Q_{\text{cl}}}\left[\frac{\tilde{\omega}_{\text{drive}}^2}{Q_{\text{cl}}\left(1-\tilde{\omega}_{\text{drive}}^2\right)} + Q_{\text{cl}}\left(1-\tilde{\omega}_{\text{drive}}^2\right)\right], \quad \text{for } \tilde{\omega}_{\text{drive}} \neq 1. \tag{13}$$

Therefore, when $\Delta A_{\text{ts}}$ is sufficiently small, $k_{\text{ts}}$ can be linearly approximated as follows:

$$k_{\text{ts}} \approx \beta_{\Delta A\to k} \Delta A_{\text{ts}}, \tag{14}$$

where $\beta_{\Delta A\to k}$, which denotes the conversion coefficient from $\Delta A_{\text{ts}}$ to $k_{\text{ts}}$, is obtained as follows:

$$\begin{aligned} \beta_{\Delta A\to k} &\equiv \frac{1}{A_{\text{free}}} \lim_{\tilde{A}_{\text{cl}}\to 1} \frac{\partial k_{\text{ts}}}{\partial \tilde{A}_{\text{cl}}} \\ &= \frac{1}{A_{\text{free}}} \frac{k_{\text{cl}}}{Q_{\text{cl}}}\left[\frac{\tilde{\omega}_{\text{drive}}^2}{Q_{\text{cl}}\left(1-\tilde{\omega}_{\text{drive}}^2\right)} + Q_{\text{cl}}\left(1-\tilde{\omega}_{\text{drive}}^2\right)\right], \quad \text{for } \tilde{\omega}_{\text{drive}} \neq 1. \end{aligned} \tag{15}$$

Similarly, the equation for $\langle F_{\text{ts}} \rangle$ is expressed as follows [59]:

$$\begin{aligned} \lim_{A_{\text{free}}\to\infty} \langle F_{\text{ts}} \rangle &= \frac{k_{\text{cl}} A_{\text{free}}}{2}\left[-\left(1-\tilde{\omega}_{\text{drive}}^2\right)\tilde{A}_{\text{cl}} \overset{\text{rep}}{\underset{\text{att}}{\pm}} \sqrt{\left(1-\tilde{\omega}_{\text{drive}}^2\right)^2 + \left(\tilde{\omega}_{\text{drive}} / Q_{\text{cl}}\right)^2\left(1-\tilde{A}_{\text{cl}}^2\right)}\right] \\ &= -\frac{A_{\text{free}}\tilde{A}_{\text{cl}}}{2} \lim_{A_{\text{free}}\to 0} k_{\text{ts}}. \end{aligned} \tag{16}$$

When excited at $f_0$, this equation can be simplified to

$$\lim_{A_{\text{free}}\to\infty} \langle F_{\text{ts}}(\tilde{\omega}_{\text{drive}} = 1) \rangle = \overset{\text{rep}}{\underset{\text{att}}{\pm}} \frac{k_{\text{cl}} A_{\text{free}}}{2Q_{\text{cl}}} \sqrt{1-\tilde{A}_{\text{cl}}^2}. \tag{17}$$

When excited at the resonance slope, the linear approximation can be used as follows:

$$\langle F_{\text{ts}} \rangle \approx \alpha_{\Delta A\to F} \Delta A_{\text{ts}}, \tag{18}$$

where $\alpha_{\Delta A\to F}$ represents the conversion coefficient from $\Delta A_{\text{ts}}$ to $\langle F_{\text{ts}} \rangle$, which is expressed by

$$\begin{aligned} \alpha_{\Delta A\to F} &= -\frac{k_{\text{cl}}}{2Q_{\text{cl}}}\left[\frac{\tilde{\omega}_{\text{drive}}^2}{Q_{\text{cl}}\left(1-\tilde{\omega}_{\text{drive}}^2\right)} + Q_{\text{cl}}\left(1-\tilde{\omega}_{\text{drive}}^2\right)\right] \\ &= -\frac{A_{\text{free}}}{2} \beta_{\Delta A\to k}, \quad \text{for } \tilde{\omega}_{\text{drive}} \neq 1. \end{aligned} \tag{19}$$

Eqs. (15) and (19) are valid for $\tilde{\omega}_{\text{drive}} < 1$ in the repulsive regime and $\tilde{\omega}_{\text{drive}} > 1$ in the attractive regime.

Figures 1c and 1d show the dependence of $\beta_{\Delta A \to k}$ on $\tilde{\omega}_{\text{drive}}$ in the repulsive and attractive regimes, respectively, as obtained from Eq. (15). In the repulsive regime, $\beta_{\Delta A \to k}$ decreases with increasing $\tilde{\omega}_{\text{drive}}$, reaches a minimum, and then increases steeply near $f_0$. In contrast, the attractive regime exhibits a laterally inverted trend: $\beta_{\Delta A \to k}$ decreases with decreasing frequency, reaches a minimum, and then sharply increases near $f_0$. These results are equivalent to those obtained from $\langle F_{\text{ts}} \rangle$ (Fig. 2(d,f) in Ref. [59]), and thus, Eq. (15) was confirmed to have an identical form as Eq. (19), differing only in the coefficients.

As with $\langle F_{\text{ts}} \rangle$, by solving $\partial \beta_{\Delta A \to k} / \partial \tilde{\omega}_{\text{drive}} = 0$, the frequency at which β reaches its minimum, referred to the lower MinForce (LMF) and UMF frequencies ($f_{\text{LMF}}$ and $f_{\text{UMF}}$), as can be determined as follows (see the arrows):

$$\tilde{\omega}_{\text{LMF}} \equiv \frac{f_{\text{LMF}}}{f_0} = \sqrt{1 - \frac{1}{Q_{\text{cl}}}}, \quad \text{for } Q_{\text{cl}} \geq 1, \tag{20}$$

$$\tilde{\omega}_{\text{UMF}} \equiv \frac{f_{\text{UMF}}}{f_0} = \sqrt{1 + \frac{1}{Q_{\text{cl}}}}. \tag{21}$$

Consequently, by exciting at the MinForce frequencies, both the sensitivities of $\langle F_{\text{ts}} \rangle$ and $k_{\text{ts}}$ can be maximized. By substituting these equations into Eq. (14), a more straightforward analytical solution for $k_{\text{ts}}$ at $f_{\text{LMF}}$ and $f_{\text{UMF}}$ can be derived as follows:

$$k_{\text{ts}}(\tilde{\omega}_{\text{drive}} = \tilde{\omega}_{\text{LMF}}) \approx \frac{k_{\text{cl}}}{Q_{\text{cl}}} \left( 2 - \frac{1}{Q_{\text{cl}}} \right) \frac{\Delta A_{\text{ts}}}{A_{\text{free}}}, \quad \text{for } \langle F_{\text{ts}} \rangle > 0, Q_{\text{cl}} \geq 1, \tag{22}$$

$$k_{\text{ts}}(\tilde{\omega}_{\text{drive}} = \tilde{\omega}_{\text{UMF}}) \approx -\frac{k_{\text{cl}}}{Q_{\text{cl}}} \left( 2 + \frac{1}{Q_{\text{cl}}} \right) \frac{\Delta A_{\text{ts}}}{A_{\text{free}}}, \quad \text{for } \langle F_{\text{ts}} \rangle < 0. \tag{23}$$

Furthermore, even when driving at $f_{\text{peak}}$, a linear approximation is valid only if both $Q_{\text{cl}}$ and $\Delta A_{\text{ts}}$ are sufficiently small. In this case, by substituting Eq. (11) into Eq. (14), the following analytical solution can be obtained.

$$k_{\text{ts}}(\tilde{\omega}_{\text{drive}} = \tilde{\omega}_{\text{peak}}) \approx 2k_{\text{cl}}\left(1 - \frac{1}{4Q_{\text{cl}}^2}\right)\frac{\Delta A_{\text{ts}}}{A_{\text{free}}}, \quad \text{for } \langle F_{\text{ts}} \rangle > 0, \tag{24}$$

The equations derived here will be used to derive the equation for MDF in Section 9 and subsequent sections.

## 3. AM noise at resonance frequency

To derive the expression for MDF in AM-AFM, we must understand the therory of AM noise. Among the various noise sources, the thermal Brownian noise of the cantilever becomes dominant in state-of-the-art AFM instruments [28,58,72,73], including HS-AFM systems [63,74,75], owing to improvements of the optical beam deflection (OBD) system. Consequently, the effect of other noises is not considered in the present theory. In this section, we derive the AM noise expression for excitation at $f_0$, following past studies [1,5]. The cantilever displacement due to thermal noise $\left\langle z_{\mathrm{th}}^2 \right\rangle$ is given by the equipartition theorem as follows [1,2,30,56,58]:

$$\left\langle z_{\mathrm{th}}^2 \right\rangle = \frac{k_{\mathrm{B}}T}{k_{\mathrm{cl}}} = \frac{1}{2\pi}\int_0^{\infty} n_{\mathrm{th}}(\omega)^2 \mathrm{d}\omega, \tag{25}$$

where $k_{\mathrm{B}}$ and $T$ are the Boltzmann constant and system temperature, respectively. By coupling this equation with $G_{\mathrm{cl}}$, the Brownian noise spectral density in the cantilever deflection ($n_{\mathrm{th}}$) is obtained as follows [30,58,76]:

$$n_{\mathrm{th}}(\omega) = \left| G_{\mathrm{cl}}(\omega) \right| F_{\mathrm{th}}, \tag{26}$$

where the thermal force $F_{\mathrm{th}}$ is expressed by

$$F_{\mathrm{th}} = \sqrt{\frac{4k_{\mathrm{cl}}k_{\mathrm{B}}T}{Q_{\mathrm{cl}}\omega_0}}. \tag{27}$$

Therefore, we obtain the common expression of $n_{\mathrm{th}}$ as follows:

$$n_{\mathrm{th}}(\omega) = \sqrt{\frac{4k_{\mathrm{B}}T}{k_{\mathrm{cl}}Q_{\mathrm{cl}}\omega_0} \frac{1}{\left(1-\tilde{\omega}^2\right)^2 + \left(\tilde{\omega}/Q_{\mathrm{cl}}\right)^2}}. \tag{28}$$

By approximating $G_{\mathrm{cl}}$ by a Taylor polynomial near $f_0$ and imposing the condition $Q_{\mathrm{cl}} \gg 1$, it can be expressed as the transfer function of first-order lag as follows [58,76]:

$$\lim_{\omega\to\omega_0} G_{\mathrm{cl}}(\omega) \approx \frac{1}{k_{\mathrm{cl}}} \frac{-\mathrm{i}Q_{\mathrm{cl}}}{1+\mathrm{i}\left(\omega_{\mathrm{m}}/\omega_{\mathrm{cl}}\right)}, \qquad (29)$$
$$\omega_{\mathrm{m}} = \omega - \omega_0,$$

where $\omega_{\mathrm{cl}}$ is the radial bandwidth of the cantilever expressed by

$$\omega_{\mathrm{cl}} = \frac{\omega_0}{2Q_{\mathrm{cl}}} = 2\pi f_{\mathrm{cl}}. \qquad (30)$$

Using this equation, $n_{\mathrm{th}}$ near $f_0$ can be approximately calculated as follows [58,76]:

$$\lim_{\omega\to\omega_0} n_{\mathrm{th}}(\omega) \approx \sqrt{\frac{4k_{\mathrm{B}}TQ_{\mathrm{cl}}}{k_{\mathrm{cl}}\omega_0} \frac{1}{1+\left(\omega_{\mathrm{m}}/\omega_{\mathrm{cl}}\right)^2}}. \qquad (31)$$

The AM noise density $n_{\mathrm{th}}^{\mathrm{AM}}$ is obtained by summing the noise spectral densities over the upper and lower sidebands as follows:

$$n_{\mathrm{th}}^{\mathrm{AM}}(\tilde{\omega}_{\mathrm{drive}}=1, \omega_{\mathrm{AM}}) = \sqrt{2}n_{\mathrm{th}}(\omega) = \sqrt{\frac{4k_{\mathrm{B}}TQ_{\mathrm{cl}}}{k_{\mathrm{cl}}\omega_0} \frac{2}{1+\left(\omega_{\mathrm{AM}}/\omega_{\mathrm{cl}}\right)^2}}. \qquad (32)$$

Thus, the total AM noise $N_{\mathrm{th}}^{\mathrm{AM}}$ is obtained by integrating $n_{\mathrm{th}}^{\mathrm{AM}}$ as follows:

$$\begin{aligned} N_{\mathrm{th}}^{\mathrm{AM}}\left(\tilde{\omega}_{\mathrm{drive}}=1\right) &= \sqrt{\frac{1}{2\pi}\int_0^{2\pi B_{\mathrm{ENBW}}} n_{\mathrm{th}}^{\mathrm{AM}}(\tilde{\omega}_{\mathrm{drive}}=1, \omega_{\mathrm{AM}})^2 \mathrm{d}\omega_{\mathrm{AM}}} \\ &= \sqrt{\frac{4k_{\mathrm{B}}TQ_{\mathrm{cl}}}{k_{\mathrm{cl}}\omega_0} 2f_{\mathrm{cl}} \tan^{-1}\left(B_{\mathrm{ENBW}}/f_{\mathrm{cl}}\right)}, \end{aligned} \qquad (33)$$

where $B_{\mathrm{ENBW}}$ represents the equivalent noise bandwidth (ENBW), whose detail will be described in Section 5. Since the relationship $\omega_{\mathrm{cl}} > 2\pi B_{\mathrm{ENBW}}$ always holds, this equation can be approximated by

$$N_{\mathrm{th}}^{\mathrm{AM}}\left(\tilde{\omega}_{\mathrm{drive}}=1\right) \approx \sqrt{\frac{4k_{\mathrm{B}}TQ_{\mathrm{cl}}}{k_{\mathrm{cl}}\omega_0}\left(2B_{\mathrm{ENBW}}\right)}. \qquad (34)$$

This equation was first derived in the early stages of AM-AFM development [2]; however, a factor of $\sqrt{2}$ was not included. This factor is necessary to account for two sidebands, as already incorporated in recent studies [5,56].

## 4. AM noise at arbitrary driving frequencies

We next consider the dependence of the AM noise on $f_{\mathrm{drive}}$, the characteristics of which remain unknown. Due to the difficulty of analytically calculating the noise characteristics, numerical simulations were performed with a time interval of 10 ns, corresponding to a sampling frequency of 100 MHz (Fig. 2(a)). Initially, Gaussian noise generated using the Box–Muller algorithm was superimposed on the cantilever drive signal. Subsequently, fast Fourier transform (FFT) was performed to convert the time-domain signal into the frequency domain, and the noise spectrum was generated by multiplying it with the transfer function of the cantilever. The frequency-domain signal was then converted back into the time domain via an inverse FFT, followed by amplitude detector processing to calculate an amplitude signal. Finally, an FFT was performed again to transform it back into the frequency domain, generating the AM noise spectrum.

For an amplitude detector, we implemented a Fourier-analysis-based (FAB) algorithm (Fig. 2(b)), which is commonly used in HS-AFM setups [28,77]. In this algorithm, the deflection signal is separately multiplied by the in-phase and quadrature components, and after integration, the Root-Sum-of-Squares (RSS) is calculated to obtain the amplitude signal. Replacing the integration with an LPF results in an implementation equivalent to a lock-in amplifier [56], which is widely used in standard AM-AFM instruments. The transfer function of the FAB in the time domain $A_{\mathrm{AM}}^{\mathrm{FAB}}$ is expressed as the convolution integral of the instantaneous amplitude $A_{\mathrm{AM}}$ and an impulsive response of the detector $f_{\mathrm{det}}$ as follows:

$$A_{\mathrm{AM}}^{\mathrm{FAB}}(t)=\int_{-\infty}^{\infty} A_{\mathrm{AM}}(t-\tau) f_{\mathrm{det}}(\tau)\,\mathrm{d}\tau, \tag{35}$$

where $f_{\mathrm{det}}$ is expressed as a rectangular function with an integral time $\Delta T_{\mathrm{integ}}$ as follows:

$$f_{\mathrm{det}}(t)=\begin{cases}\dfrac{1}{\Delta T_{\mathrm{integ}}}, & \text{for } 0\le t<\Delta T_{\mathrm{integ}},\\ 0 & \text{elsewhere,}\end{cases} \tag{36}$$

$$\Delta T_{\mathrm{integ}}=\frac{2\pi n_{\mathrm{integ}}}{\omega_{\mathrm{drive}}}, \tag{37}$$

where $n_{\mathrm{integ}}$ is the integration number of the cantilever oscillation period. By performing a Fourier transform, Eq. (35) can be converted to a cross correlation as follows:

$$\hat{A}_{\mathrm{AM}}^{\mathrm{FAB}}(\mathrm{i}\omega_{\mathrm{AM}})=\hat{A}_{\mathrm{AM}}(\mathrm{i}\omega_{\mathrm{AM}})\hat{f}_{\mathrm{det}}(\mathrm{i}\omega_{\mathrm{AM}}), \tag{38}$$

where $\hat{A}_{\mathrm{AM}}^{\mathrm{FAB}}$, $\hat{A}_{\mathrm{AM}}$, and $\hat{f}_{\mathrm{det}}$ represent the Fourier transform of $A_{\mathrm{AM}}^{\mathrm{FAB}}$, $A_{\mathrm{AM}}$, and $f_{\mathrm{det}}$, respectively, and $\hat{f}_{\mathrm{det}}$ is expressed as

$$\begin{aligned}\hat{f}_{\mathrm{det}}(\mathrm{i}\omega_{\mathrm{AM}})&=\int_{-\infty}^{\infty}f_{\mathrm{det}}(t)\exp\left(-\mathrm{i}\omega_{\mathrm{AM}}t\right)\mathrm{d}t\\&=\frac{1-\exp\left(-\mathrm{i}\Delta T_{\mathrm{integ}}\omega_{\mathrm{AM}}\right)}{\mathrm{i}\Delta T_{\mathrm{integ}}\omega_{\mathrm{AM}}}.\end{aligned} \tag{39}$$

The absolute value can be expressed as a sinc function, as follows:

$$\left|\hat{f}_{\mathrm{det}}(\omega_{\mathrm{AM}})\right|=\left|\mathrm{sinc}\left(\frac{\Delta T_{\mathrm{integ}}\omega_{\mathrm{AM}}}{2}\right)\right|. \tag{40}$$

We performed simulations with common HS-AFM experimental parameters: $k_{\mathrm{cl}}$ = 0.1 N/m, $f_0$ = 1 MHz, and $T$ = 298 K. The data presented in each noise spectrum result from 400 integrations. In Fig. 2(c,d), as theoretically predicted, the simulated spectra exhibit the sinc-function characteristic, with flat response at low frequencies but a sharp drop at multiples of $f_{\mathrm{drive}}$. Furthermore, the cutoff frequency tends to be reduced to lower frequencies concurrently with increasing $n_{\mathrm{integ}}$. However, in contrast to $Q_{\mathrm{cl}}$ = 1.5, no significant change was observed for $Q_{\mathrm{cl}}$ = 10. To examine this further, $n_{\mathrm{th}}^{\mathrm{AM}}$ (Eq. (32)) is also plotted in Fig. 2(c,d) (yellow broken curves). The results demonstrate that for $Q_{\mathrm{cl}}$ = 1.5, the actual bandwidth is less than that of the cantilever, indicating that the amplitude detector limits the total bandwidth. Conversely, for $Q_{\mathrm{cl}}$ = 10, no notable discrepancy is observed, suggesting that the cantilever's bandwidth is the predominant limiting factor.

We next examine the AM noise spectra with varying $f_{\mathrm{drive}}$ from 0.5 to 1.5 MHz. Although the excitation efficiency changes depending on $f_{\mathrm{drive}}$, simulations were performed while adjusting the excitation power to maintain a constant $A_{\mathrm{free}}$ across all $f_{\mathrm{drive}}$. In Fig. 2(e,f), the noise level was predominantly consistent at low frequencies but exhibited a gradual decline at higher frequencies for $f_{\mathrm{drive}}$ values of 0.5–1 MHz. It is also evident that the noise level at low frequency varies depending on $f_{\mathrm{drive}}$, peaking at $f_{\mathrm{drive}} = f_0$. This trend is more pronounced for $Q_{\mathrm{cl}}$ = 10 than for $Q_{\mathrm{cl}}$ = 1.5, suggesting that the noise level can be reduced by excitation at a frequency other than $f_0$. However, at $f_{\mathrm{drive}}$ = 1.5 MHz, a gain noise peak appeared in the high frequency range (indicated by the arrow), which may pose to a risk of feedback oscillation.

To quantify the $f_{\mathrm{drive}}$ dependence of the AM noise density, the average low-frequency AM noise was calculated from ~100 integrated spectra, as depicted in the bottom panel of Fig. 2(a). In Fig. 3(a,b), we first investigate the $A_{\mathrm{free}}$ dependence. When excited at $A_{\mathrm{free}}$ = 1 $\mathrm{nm_{p\text{–}0}}$ or higher, the results for both $Q_{\mathrm{cl}}$ = 1.5 and $Q_{\mathrm{cl}}$ = 10 match the SHO approximation as follows:

$$n_{\mathrm{th}}^{\mathrm{AM}}(\tilde{\omega}_{\mathrm{drive}}) = \sqrt{\frac{4k_B T Q_{\mathrm{cl}}}{\omega_0 k_{\mathrm{cl}}} \frac{2}{\left[Q_{\mathrm{cl}}\left(1-\tilde{\omega}_{\mathrm{drive}}^2\right)\right]^2 + \tilde{\omega}_{\mathrm{drive}}^2}}, \quad \text{for } \omega_{\mathrm{AM}} \ll \omega_{\mathrm{cl}}, \tag{41}$$

where the factor of $\sqrt{2}$ is included to account for the two sidebands. However, reducing $A_{\mathrm{free}}$ below 1 $\mathrm{nm_{p\text{–}0}}$ causes the noise levels near the resonance peak to decrease, while those outside the peak increase. Consequently, the peak shape gradually flattens, and at $A_{\mathrm{free}}$ = 0 nm, the peak noise is reduced to about half.

To investigate the cause of the non-ideal noise characteristics at excessively low amplitudes, we analyzed the output of the cosine integral, bypassing the RSS calculation in the FAB (cosine output in Fig. 2(b)). In Fig. 3(c,d), the results perfectly matched the SHO model, even when the drive amplitude is zero. This finding indicates that this phenomenon arises from the coupling of high-frequency and low-frequency noises due to the RSS calculation, which occurs only when Eqs.

(42) and (43) are not satisfied.

We also examined the $n_{\text{integ}}$ dependence of the noise spectra (Fig. 3(e,f)). With excitation at $A_{\text{free}}$ = 0.3 $\text{nm}_{\text{p-0}}$, $n_{\text{integ}}$ was varied from 1 to 8, and the results asymptotically approached the SHO model as $n_{\text{integ}}$ increases. For $Q_{\text{cl}}$ = 1.5, results matched the SHO model at $n_{\text{integ}} \geq 8$. For $Q_{\text{cl}}$ = 10, the SHO model was achieved at $n_{\text{integ}} \geq 32$. This result implies that, even if $A_{\text{free}}$ falls below the noise criterion, reducing the bandwidth of the amplitude detector can suppress the coupling.

To summarize, the criterion for the SHO model validity was found to be determined by SNR regardless of $Q_{\text{cl}}$. By substituting the simulation parameter $k_{\text{cl}}$ = 0.1 N/m into Eq. (25), the RMS noise $\sqrt{\langle z_{\text{th}}^2 \rangle}$ is calculated to be 0.2 nm at 298 K. Thus, by comparing with $A_{\text{cl}}$ = 1 $\text{nm}_{\text{p-0}}$, we obtain the empirical SHO criterion equation as follows:

$$A_{\text{free}} \geq 5\sqrt{\langle z_{\text{th}}^2 \rangle}. \tag{42}$$

By coupling with Eq. (25), we obtain the more practical equation as follows:

$$A_{\text{free}}\sqrt{k_{\text{cl}}} \geq 5\sqrt{k_{\text{B}}T} \approx 0.32\ \text{nm}\sqrt{\text{N/m}}, \quad \text{for } T = 298\ \text{K}. \tag{43}$$

In ambient conditions, large values of $A_{\text{free}}\sqrt{k_{\text{cl}}}$ (typically $\geq 9\ \text{nm}\sqrt{\text{N/m}}$; e.g., $A_{\text{free}}$ = 5.2 $\text{nm}_{\text{p-0}}$ and $k_{\text{cl}}$ = 3 N/m) are used to avoid jump-to-contact. Although this effect is mitigated in liquid, values of $A_{\text{free}}\sqrt{k_{\text{cl}}}$ greater than $1\ \text{nm}\sqrt{\text{N/m}}$ (e.g., $A_{\text{free}}$ = 3 $\text{nm}_{\text{p-0}}$ and $k_{\text{cl}}$ = 0.1 N/m, which are typical parameters for HS-AFM) are still required for stable imaging. Thus, this SHO criterion is sufficiently satisfied in most measurement systems. Therefore, we obtain the total AM noise expression that accounts for $\tilde{\omega}_{\text{drive}}$ as follows:

$$N_{\text{th}}^{\text{AM}}(\tilde{\omega}_{\text{drive}}) \approx \sqrt{\frac{4k_{\text{B}}TQ_{\text{cl}}}{k_{\text{cl}}\omega_0} \frac{(2B_{\text{ENBW}})}{\left[Q_{\text{cl}}\left(1-\tilde{\omega}_{\text{drive}}^2\right)\right]^2 + \tilde{\omega}_{\text{drive}}^2}}. \tag{44}$$

## 5. Estimation of equivalent noise bandwidth

This section describes the determination of $B_{\mathrm{ENBW}}$ required to calculate MDF. Since the total AM noise significantly varies as a function of $B_{\mathrm{ENBW}}$, a clear understanding of how to calculate $B_{\mathrm{ENBW}}$ is essential, yet this has rarely been explained in previous studies. The $B_{\mathrm{ENBW}}$ represents the bandwidth assumed for an ideal brick-wall filter and can be obtained by integrating the entire power spectral density [78]. The $B_{\mathrm{ENBW}}$ is not an intrinsic instrument parameter but rather a user-adjustable parameter limited by the total bandwidth of the cantilever and amplitude detector. Furthermore, the $B_{\mathrm{ENBW}}$ depends on how the MDF is evaluated, as described below.

To evaluate the MDF of AFM system used as a force sensor or for force curve measurements, the $B_{\mathrm{ENBW}}$ can be simply determined from the power spectral density of the amplitude signal. In high-$Q_{\mathrm{cl}}$ conditions, the system can be approximated using $G_{\mathrm{cl}}$ alone since the cantilever limits the total bandwidth. As shown in Fig. 2(d), when $f_{\mathrm{drive}}$ is set near the resonance peak—as is typically the case in actual measurements—the system can be approximated as a first-order lag system, as expressed in Eq. (29). Thus, $B_{\mathrm{ENBW}}$ is derived as follows:

$$\begin{aligned} B_{\mathrm{ENBW}}^{\mathrm{1st}} &= \int_0^{\infty} \left|G_{\mathrm{cl}}(f)\right|^2 \mathrm{d}f = \int_0^{\infty} \frac{1}{1+\left(f/f_{\mathrm{cl}}\right)^2} \mathrm{d}f \\ &= \frac{\pi}{2} f_{\mathrm{cl}} \approx 1.57 f_{\mathrm{cl}}. \end{aligned} \tag{45}$$

This indicates that the $B_{\mathrm{ENBW}}$ of the cantilever is larger than $f_{\mathrm{cl}}$. By substituting Eq. (30), a more explicit expression can be obtained as follows:

$$B_{\mathrm{ENBW}}^{\mathrm{1st}} = \frac{\pi f_0}{4Q_{\mathrm{cl}}}. \tag{46}$$

Under low-$Q_{\mathrm{cl}}$ conditions, when using a lock-in amplifier, its LPF typically limits the total bandwidth. Since lock-in amplifiers generally employ a Butterworth-type LPF, the $B_{\mathrm{ENBW}}$ for the second- and fourth-order cases is calculated as follows：

$$\begin{cases} B_{\mathrm{ENBW}}^{\mathrm{2nd}} = \int_0^{\infty} \frac{1}{1+\left(f/f_{\mathrm{c}}\right)^4} \mathrm{d}f = \frac{\pi}{2\sqrt{2}} f_{\mathrm{c}} \approx 1.11 f_{\mathrm{c}}, \\ B_{\mathrm{ENBW}}^{\mathrm{4th}} = \int_0^{\infty} \frac{1}{1+\left(f/f_{\mathrm{c}}\right)^8} \mathrm{d}f = \frac{\pi f_{\mathrm{c}}}{8\sin\left(\pi/8\right)} \approx 1.03 f_{\mathrm{c}}, \end{cases} \tag{47}$$

where $f_{\mathrm{c}}$ denotes the cutoff frequency. This indicates that, in practice $B_{\mathrm{ENBW}}$ is almost same as $f_{\mathrm{c}}$.

Meanwhile, to evaluate the MDF during topographic imaging, the effects of the PID controller and Z-scanner must be considered. The integrator in the PID controller accumulates the error signal over time, enabling AFM imaging to be performed even when the SNR of the amplitude detector output is less than 1 (i.e., $\Delta A_{\mathrm{ts}}$ is smaller than the noise level). However, when the integration time becomes longer, the bandwidth becomes more limited, resulting in a reduction in $B_{\mathrm{ENBW}}$. Furthermore, since the Z-scanner acts as a second-order lag system, its transfer function must also be taken into account.

In AFM, the frequency at which the phase delay reaches 45° is commonly used to evaluate the feedback bandwidth, referred to as $f_{45^\circ}$. The expression of $f_{45^\circ}$ can be obtained by linearly approximating the phase characteristics at low frequencies, as follows [77,79]:

$$f_{45^\circ} = \left( \underbrace{\frac{8}{\pi} \frac{Q_{\mathrm{cl}}}{f_0}}_{\text{cantilever}} + \underbrace{\frac{4}{\pi} \frac{\Delta\varphi_{\mathrm{amp}}}{f_{\mathrm{drive}}} + \frac{8}{\pi} \frac{Q_z}{f_z} + 8\delta}_{\text{other}} \right)^{-1}, \tag{48}$$

where $\Delta\varphi_{\mathrm{amp}}$, $Q_z$, $f_z$, and $\delta$ denote the phase delay in the amplitude detector, Q-factor and resonance frequency of the Z-scanner, and the latency in the electronic circuits, respectively. The first term represents the latency due to the cantilever, while the subsequent terms correspond to contributions from other components. As described in Section 2, since $f_{\mathrm{drive}}$ generally does not coincide with $f_0$, we slightly modified the original equation by replacing $f_0$ with $f_{\mathrm{drive}}$ in the parameter of the amplitude detector in the second term.

Since the cantilever and the integral controller each exhibit first-order roll-off, and the Z-scanner exhibits second-order roll-off, the overall system exhibits roll-off characteristics ranging from first to fourth order, depending on which component limits the bandwidth. In particular, when the cantilever is not the bandwidth-limiting component, the system exhibits a second- to fourth-order roll-off, and $B_{\text{ENBW}}$ is expressed as follows (Supplementary Note 1):

$$\begin{cases} B_{\text{ENBW}}^{\text{2nd}} = \dfrac{\pi Q_{\text{total}}^2}{-1+\sqrt{\left(2Q_{\text{total}}\right)^2+1}} f_{45^\circ}, \\ B_{\text{ENBW}}^{\text{4th}} = \dfrac{\pi}{2} \dfrac{Q_{\text{total}}^2\left(Q_{\text{total}}^2+1\right)\left(\sqrt{2}-1\right)}{-1+\sqrt{\left[2Q_{\text{total}}\left(\sqrt{2}-1\right)\right]^2+1}} f_{45^\circ}, \end{cases} \tag{49}$$

where $Q_{\text{total}}$ denotes the overall Q-factor of the system. Assuming it is in the range of 0.5 to 0.7, the coefficients for the second- and fourth-order cases are estimated to be approximately 1.9–2.1 and 2.5–3.0, respectively. Since these coefficients depend on both the experimental setup and imaging parameters, they should be chosen within the range of 1.9 to 3.

By coupling Eqs. (48) and (49), $B_{\text{ENBW}}$ can be determined. However, since Eq. (48) is based on a linear approximation, it fails to match the exact solution of Eq. (30) in the limit of infinite $Q_{\text{cl}}$, where the cantilever bandwidth becomes the rate-limiting factor. Meanwhile, we found that setting the coefficient to 2 allows the $B_{\text{ENBW}}$ to match the exact solution of Eq. (46) even in the limit of infinite $Q_{\text{cl}}$, as follows:

$$B_{\text{ENBW}} = 2\left(\underbrace{\frac{8}{\pi}\frac{Q_{\text{cl}}}{f_0}}_{\text{cantilever}} + \underbrace{\frac{4}{\pi}\frac{\Delta\varphi_{\text{amp}}}{f_{\text{drive}}} + \frac{8}{\pi}\frac{Q_z}{f_z}}_{\text{other}}\right)^{-1} = 2\left(\frac{8}{\pi}\frac{Q_{\text{cl}}}{f_0} + \delta_{\text{amp,z}}\right)^{-1}, \tag{50}$$

Therefore, this equation enables the calculation of $B_{\text{ENBW}}$ without discontinuities for any value of $Q_{\text{cl}}$. Since $\delta$ typically arises from only the phase delay, it is not considered in this equation.

We next examine the $Q_{\text{cl}}$-dependence of $B_{\text{ENBW}}$ using Eq. (50). In a typical HS-AFM setup with a setpoint ratio of 0.9, $f_{45^\circ}$ ranges from 10 to 30 kHz [28]. Therefore, we assumed $B_{\text{ENBW}}$ at $Q_{\text{cl}} = 1.5$ to be 60 kHz and the latency other than that of the cantilever ($\delta_{\text{amp,z}}$) to 29.5 µs to satisfy this condition. These parameters are also used in the calculations presented in subsequent sections.

In Fig. 4(a), when $Q_{\text{cl}}$ is low, $\delta_{\text{amp,z}}$ is the rate-limiting factor, resulting in a nearly flat $B_{\text{ENBW}}$. As $Q_{\text{cl}}$ increases, $B_{\text{ENBW}}$ gradually approaches the cantilever-limited value described by Eq. (46). We also analyzed a conventional AM-AFM system. Most commercial AFMs employ piezoelectric tube actuators, whose $f_z$ typically range from 2 to 8 kHz [80-82], with $Q_z \approx 10$. Based on these parameters, $\delta_{\mathrm{amp,z}}$ is set to 5000 μs. In this case, the Z-scanner bandwidth limits $B_{\mathrm{ENBW}}$, resulting in a nearly flat response across the entire $Q_{\mathrm{cl}}$ range. Furthermore, we examine a condition where $Q_z$ is reduced to 0.7 using a $Q_z$ controller [28], and $\delta_{\mathrm{amp,z}}$ is set to 360 μs. Although this led to an overall increase in $B_{\mathrm{ENBW}}$, it remains nearly flat over most $Q_{\mathrm{cl}}$ values.

The saturation of $B_{\mathrm{ENBW}}$ observed when using a conventional Z scanner is attributed to the use of a cantilever with $f_0$ = 1 MHz, which is commonly used for HS-AFM. Therefore, we also consider a cantilever with $f_0$ = 100 kHz, which is more commonly used in AM-AFM. In Fig. 4(b), the results indicate a reduction in the cantilever's $B_{\mathrm{ENBW}}$, leading to a suppression of the overall $B_{\mathrm{ENBW}}$ in the high-$Q_{\mathrm{cl}}$ values. These findings suggest that when $Q_{\mathrm{cl}}$ is low, $B_{\mathrm{ENBW}}$ is limited by $\delta_{\mathrm{amp,z}}$ and remains nearly constant, whereas at higher $Q_{\mathrm{cl}}$ values, it asymptotically approaches the value determined by the cantilever latency.

## 6. Experimental deviations from SHO model at Low $Q_{cl}$

In the previous sections, we formulated the theories based on the SHO model. To verify its applicability to real experiments, we performed experimental validation using a HS-AFM setup, described in our previous study [59], with a small $Si_3N_4$ rectangular cantilever (BLAC10DS-A2, Olympus). Since the OBD method was employed, a dynamic-to-static correction factor ($\chi$) was applied to the optical lever sensitivity [63,71]. We assumed $\chi$ = 0.97 because the cantilever and the laser spot were comparable in size, the spot was placed near the center.

Figure 5(a) shows the experimental noise spectrum, which was fitted to the practical theoretical noise spectrum ($n_{th,fitted}$) that accounts for the noise floor $n_{floor}$, as shown below:

$$n_{th,fitted}(\omega) = \sqrt{n_{th}(\omega)^2 + n_{floor}^2}. \quad (51)$$

Two fitting cases are shown in Fig. 5(a): one across the full frequency range, including the low-frequency region ($Q_{cl}$ = 1.82, $k_{cl}$ = 0.153 N/m), and another only near the resonance peak ($Q_{cl}$ = 1.68, $k_{cl}$ = 0.142 N/m). The following parameters were shared between the two cases: $f_0$ = 540 kHz and $n_{floor}$ = 40 fm/√Hz. In the former case, the fit agreed reasonably well with the entire spectrum, though the experimental peak was slightly broader. In the latter case, although the low-frequency side of the peak fitted slightly better, the off-resonance gain was overestimated. In both cases, discrepancies were evident on the high-frequency side of the peak (see the arrow in Fig. 5(a)). Thus, although both fitting methods are commonly used in practical experiments, perfect fitting could not be achieved regardless of the choice of fitting parameters.

This discrepancy is likely due to the fact that the SHO model no longer holds strictly when the $Q_{cl}$ becomes very low. Over more than a decade, analytical theories describing the frequency response of an explicit cantilever beam immersed in a viscous fluid have been developed by coupling the Euler–Bernoulli beam equation with the Navier–Stokes equations [64-69]. To the best of our knowledge, the most recent theory was published in 2010 by Clark et al. [69].

Hence, we also fitted the experimental results using this model, which is referred to as the Hydro-Beam (HB) model (Supplementary Note 2). The noise spectral density in this model is denoted as $n_{\mathrm{th},z}^{\mathrm{HB}}$. The fitting requires the cantilever parameters $L_{\mathrm{cl}}$, $b_{\mathrm{cl}}$, $h_{\mathrm{cl}}$, $\rho_{\mathrm{cl}}$, and $E_{\mathrm{cl}}$, which denote the length, width, thickness, density, and Young's modulus of the cantilever beam, respectively. It also requires the liquid parameters $\rho_{\mathrm{liq}}$ and $\eta_{\mathrm{liq}}$, which denote the density and viscosity of the fluid, respectively. In the experiments, higher-order resonances were hardly visible since the laser spot was placed at the center of the cantilever, where a node of the second resonance mode is located.

Since direct fitting using the original equation did not yield satisfactory results regardless of the parameter values, we introduce a gain correction factor $C_{\mathrm{gain}}$.

$$n_{\mathrm{th,fitted}}^{\mathrm{HB}}(\omega) = \sqrt{\left(C_{\mathrm{gain}} n_{\mathrm{th},z}^{\mathrm{HB}}(\omega)\right)^2 + n_{\mathrm{floor}}^2}. \tag{52}$$

This adjustment is likely necessary because the theory neglects various non-ideal factors present in the actual experiments [65].

Using this modified equation, we performed fitting with the following parameters. We set $L_{\mathrm{cl}}$ = 7.9 μm, $b_{\mathrm{cl}}$ = 1.45 μm, and $h_{\mathrm{cl}}$ = 0.090 μm, which are close to the nominal dimensions of the BLAC10DS-A2 cantilever ($L_{\mathrm{cl}}$ = 9 μm, $b_{\mathrm{cl}}$ = 2 μm, and $h_{\mathrm{cl}}$ = 0.11 μm, thickness excluding the gold reflex coating), and adopted $\rho_{\mathrm{cl}}$ = 3170 kg/m$^3$ and $E_{\mathrm{cl}}$ = 280 GPa for the $Si_3N_4$ material. To represent water, we used $\rho_{\mathrm{liq}}$ = 997 kg/m$^3$ and $\eta_{\mathrm{liq}}$ = 0.89 mPa·s at 298 K. Under these conditions, fitting was performed using $C_{\mathrm{gain}}$ = 1.15 and $n_{\mathrm{floor}}$ = 40 fm/√Hz.

As shown by the blue dashed curve in Fig. 5(b), the HB model successfully reproduced the experimental broad resonance peak. However, in the low-frequency range (see the arrow), the model slightly underestimated the noise density compared to the experimental spectrum. This discrepancy is also evident in the original paper through comparisons with numerical simulations; however, the authors deem the numerical results less reliable [69]. Thus, this may be attributed to an effects not accounted for in the model, such as the reflective gold coating [83]. Furthermore, given the simplicity of the SHO-based formulation, it is more practical to use the SHO model to define the relationships between force and amplitude.

Thus, we next evaluate how deviations from the SHO model affect force estimation accuracy. By analytically solving the Euler–Bernoulli beam equation under certain boundary conditions, $k_{\mathrm{cl}}$ can be calculated as follows:

$$k_{\mathrm{cl}} = \xi \frac{3E_{\mathrm{cl}} I_{\mathrm{cl}}}{L_{\mathrm{cl}}^{3}}, \tag{53}$$

where $\xi$ denotes the dynamic-to-static ratio of the spring constants, which is ~1.03 for the rectangular cantilever [71]. When $C_{\mathrm{gain}}$ is taken into account, $k_{\mathrm{cl}}$ derived from the HB model ($k_{\mathrm{cl}}^{\mathrm{HB}}$) is given by

$$k_{\mathrm{cl}}^{\mathrm{HB}} = \frac{k_{\mathrm{cl}}}{C_{\mathrm{gain}}^{2}}. \tag{54}$$

Using this equation, the value of $k_{\mathrm{cl}}^{\mathrm{HB}}$ was estimated to be 0.116 N/m.

As expressed in Eq. (25), the integral of the thermal noise spectrum is inversely proportional to $k_{\mathrm{cl}}$, which can lead to overestimation of $k_{\mathrm{cl}}^{\mathrm{HB}}$ due to the low-frequency deviations. To address this, we extend the HB model by adding a low-frequency term as follows:

$$n_{\mathrm{th,fitted}}^{\mathrm{mod}}(x,\omega) = \sqrt{n_{\mathrm{th,fitted}}^{\mathrm{HB}}(x,\omega)^{2} + n_{\mathrm{LF}}(\omega)^{2}}. \tag{55}$$

Here, the low-frequency component $n_{\mathrm{LF}}$ is modeled using an exponential function defined as follows:

$$n_{\mathrm{LF}}(\omega) = \alpha_{\mathrm{LF}} \exp(-\omega / \omega_{\mathrm{LF}}), \tag{56}$$

where $\alpha_{\mathrm{LF}}$ and $\omega_{\mathrm{LF}}$ are fitting parameters representing the DC noise level and cutoff frequency, respectively. Using the modified model, we obtained results that show good agreement with the experimental data across the entire frequency range, including the low-frequency region, as shown by the green dashed curve in Fig. 5(b). Furthermore, when this expression is used, the corrected $k_{\mathrm{cl}}$ ($k_{\mathrm{cl}}^{\mathrm{corr}}$) is expressed as follows:

$$k_{\mathrm{cl}}^{\mathrm{corr}} = \frac{1}{\dfrac{1}{k_{\mathrm{cl}}^{\mathrm{HB}}} + \dfrac{\alpha_{\mathrm{LF}}^2}{2k_{\mathrm{B}}T}\dfrac{\omega_{\mathrm{LF}}}{2\pi}}. \tag{57}$$

Using this equation yields more accurate results. Applying this method yielded $k_{\mathrm{cl}}^{\mathrm{corr}}$ of 0.114 N/m, with $\alpha_{\mathrm{LF}}$ = 95 fm/√Hz, $\omega_{\mathrm{LF}}$ = 2π × 150 kHz. Other parameters are identical with those used in the estimation of $k_{\mathrm{cl}}^{\mathrm{HB}}$.

Taking $k_{\mathrm{cl}}^{\mathrm{corr}}$ as a reference, we assess the accuracy of other estimation methods as below. The original HB model showed only a 2% (0.116 N/m) difference, indicating that deviations in the low-frequency region have minimal impact. In contrast, the SHO model resulted in errors of 33% (0.152 N/m) and 24% (0.142 N/m) for the full and peak fittings, respectively. This suggests that deviation in the high-frequency region significantly affects the result, and correction is necessary for accurate force measurement.

Another concern is the reduction in the magnitude of $\Delta A_{\mathrm{ts}}$, which arises from two mechanisms: (1) gain changes near the resonance peak due to variations in $f_0$ and $Q_{\mathrm{cl}}$ (hereafter "peak-change contribution"), and (2) reduced excitation efficiency (hereafter "excitation-efficiency contribution") [59]. The peak broadening seen in experiments affects only the former. Since the peak-change contribution is approximately proportional to the frequency derivative of the amplitude [1], we calculated the differentiated spectrum in Fig. 5(c). Fitting using the SHO model over the entire spectrum (blue dashed curve) and near the peak alone (green dashed curve) results in maximum noise density derivative of 0.67 and 0.6 fm/√Hz/kHz, whereas the experimental value was 0.4 fm/√Hz/kHz. These differences correspond to approximately 68% (from 0.4 to 0.67) and 50% (from 0.4 to 0.6), respectively.

For comparison, the derivative of the HB model spectrum is shown in Fig. 5(d). While the original model exhibits significant discrepancies in the low-frequency region (blue dashed line), the improved model (green dashed curve) demonstrates good agreement across the entire frequency range. Therefore, if $\Delta A_{\mathrm{ts}}$ is assumed to result solely from the resonance peak shift, the $F_{\mathrm{ts}}$ estimated using the SHO model from $\Delta A_{ts}$ would be underestimated by up to 50–68% compared to the actual $F_{ts}$. However, a previous study demonstrated that, contrary to this expectation, the actual $F_{ts}$ values measured with excitation at the resonance slope are slightly lower than the SHO-predicted ones (Fig. 6 in Ref. [59]), indicating that such a large discrepancy does not occur in practice.

As described below, this can be attributed to the fact that, at low $Q_{cl}$, the peak-change contribution becomes negligible. In Eq. (19), $\alpha_{\Delta A\to F}$ represents the conversion factor from $\Delta A_{ts}$ to force, accounting for the total contribution. Thus, the total contribution to $\Delta A_{ts}$ can be obtained by taking the reciprocal of $\alpha_{\Delta A\to F}$. Furthermore, by substituting $\tilde{\omega}_{drive}=0$ into this expression, the excitation-efficiency contribution can be obtained; thus, its ratio $R_{exc}$ is obtained by normalized with the total contribution as follows:

$$R_{exc}(\tilde{\omega}_{drive}) = \frac{1/\alpha_{\Delta A\to F}(\tilde{\omega}_{drive}=0)}{1/\alpha_{\Delta A\to F}(\tilde{\omega}_{drive})} = \frac{1}{Q_{cl}}\left[\frac{\tilde{\omega}_{drive}^2}{Q_{cl}\left(1-\tilde{\omega}_{drive}^2\right)} + Q_{cl}\left(1-\tilde{\omega}_{drive}^2\right)\right]. \tag{58}$$

Therefore, the peak-change contribution ratio $R_{peak}$ is derived as follows:

$$R_{peak}(\tilde{\omega}_{drive}) = 1 - R_{exc}(\tilde{\omega}_{drive}) = 1 - \frac{1}{Q_{cl}}\left[\frac{\tilde{\omega}_{drive}^2}{Q_{cl}\left(1-\tilde{\omega}_{drive}^2\right)} + Q_{cl}\left(1-\tilde{\omega}_{drive}^2\right)\right]. \tag{59}$$

This equation is plotted in Fig. 5(e), which shows that $R_{peak}$ is zero at $\tilde{\omega}_{drive}=0$, increases with $\tilde{\omega}_{drive}$, and reaches its maximum at $f_{LMF}$. With further increases in $\tilde{\omega}_{drive}$, $R_{peak}$ decreases again, and it becomes even negative near $\tilde{\omega}_{drive}=1$, reflecting the reduced amplitude variation in the vicinity of $f_0$. Additionally, the maximum $R_{peak}$ is approximately 10% for $Q_{cl}$ = 1.5, while it increases to around 90% for $Q_{cl}$ = 20. To investigate the $Q_{cl}$-dependence of $R_{peak}$ at $f_{LMF}$, we substitute Eq. (20) into Eq. (59) as follows:

$$R_{\text{peak}}(\tilde{\omega}_{\text{drive}} = \tilde{\omega}_{\text{LMF}}) = \left(1 - \frac{1}{Q_{\text{cl}}}\right)^2 . \tag{60}$$

The result is plotted in Fig. 4(f). As shown, for $Q_{\text{cl}} = 1.5$, $R_{\text{peak}}$ remains at 11%, but exceeds 80% when $Q_{\text{cl}} > 10$. Accordingly, when $Q_{\text{cl}} = 1.5$, the 50–68% underestimation of $F_{\text{ts}}$ is suppressed to approximately 5–7%. However, since the linear approximation is invalid at $\tilde{\omega}_{\text{drive}} = 1$, the resulting error is inevitably larger than at the resonance slope.

## 7. Experimental validation of frequency dependent AM noise

In this section, we experimentally validate the theory of AM noise formulated in Section 4. We used the same experimental setup as described in Section 6, along with an acoustic excitation method and an FAB amplitude detector. FFTs were applied to transform the amplitude signal to the AM noise spectra. Fig. 6(a) shows the AM noise spectra measured with $A_{\mathrm{free}}$ = 3 $\mathrm{nm}_{\mathrm{p-0}}$ at various $f_{\mathrm{drive}}$. As with the simulations shown in Fig. 2(e,f), the low-frequency noise level significantly varied depending on $f_{\mathrm{drive}}$.

The $f_{\mathrm{drive}}$ dependence of the average low-frequency AM noise level is plotted in Fig. 6(b) as the green line with circle markers. A spectrum with a peak around 500 kHz ($\tilde{\omega}_{\mathrm{peak}}$ = 0.93) was clearly observed. The red curve shows the AM noise converted from the experimental deflection noise spectrum in Fig. 5(a) using Eq. (32), and excellent agreement was achieved. Additionally, the SHO fit result is also shown as blue dashed curve; similar to Fig. 5(a), although the experimental peak was slightly broader, the absolute values matched well. For comparison, the AM noise converted from the deflection noise using Eq. (32) without considering the $\sqrt{2}$ factor is also shown as the gray curve. As is evident, this clearly underestimates the noise level by a factor of $\sqrt{2}$ compared to the AM noise measured directly. These results experimentally confirm the necessity of the factor of $\sqrt{2}$ in Eq. (32).

## 8. MDF from previous studies

In this section, before deriving the MDF for AM-AFM, we review MDF theories in previous literatures. The minimum detectable $k_{ts}$ and $\langle F_{ts} \rangle$ are denoted as $k_{ts,min}$ and $\langle F_{ts} \rangle_{min}$, representing approximations in the small and large $A_{free}$ limits, respectively. The theoretical $k_{ts,min}$ for FM-AFM was first derived by Albrecht et al. ($k_{ts,min}^{FM}$) [30] as follows:

$$k_{ts,min}^{FM} \overset{rep}{\underset{att}{=\mp}} \frac{1}{A_{free}} \sqrt{\frac{8k_{cl}k_B T}{Q_{cl}\omega_0} B_{ENBW}} \overset{rep}{\underset{att}{=\mp}} \sqrt{\frac{4k_{cl}k_B T}{Q_{cl}\omega_0 \langle z_{free}^2 \rangle} B_{ENBW}}, \tag{61}$$

Hereafter, unlike the original literatures, signs are included in the equations to distinguish between repulsive and attractive regimes. $\sqrt{\langle z_{free}^2 \rangle}$ represents the RMS value of $A_{free}$, which can be given by:

$$\sqrt{\langle z_{free}^2 \rangle} = \frac{A_{free}}{\sqrt{2}}. \tag{62}$$

For PM-AFM, Fukuma et al. derived the equation below [31]:

$$\delta F_c \big|_{min}^{PM} \overset{rep}{\underset{att}{=\mp}} \sqrt{\frac{8k_{cl}k_B T}{Q_{cl}\omega_0} B_{ENBW}}, \tag{63}$$

where $\delta F_c$ denotes a Virial force defined as follows:

$$\delta F_c = \frac{2}{A_{free}} \langle F_{ts} z_{tip} \rangle. \tag{64}$$

Therefore, by applying the small $A_{free}$ approximation (Eq. (7)), we obtain $k_{ts,min}$ for PM-AFM ($k_{ts,min}^{PM}$) as follows:

$$k_{ts,min}^{PM} \overset{rep}{\underset{att}{=\mp}} \frac{1}{A_{free}} \sqrt{\frac{8k_{cl}k_B T}{Q_{cl}\omega_0} B_{ENBW}}. \tag{65}$$

As is evident, $k_{ts,min}^{PM}$ is found to be equivalent to $k_{ts,min}^{FM}$, indicating that Eqs. (61) and (65) represent the theoretical dynamic-mode limit. Accordingly, we define $k_{\mathrm{ts,min}}^{\mathrm{lim}}$ as the dynamic-mode limit of $k_{\mathrm{ts,min}}$ as follows:

$$k_{\mathrm{ts,min}}^{\mathrm{lim}} \equiv k_{\mathrm{ts,min}}^{\mathrm{FM}} = k_{\mathrm{ts,min}}^{\mathrm{PM}} = \underset{\mathrm{att}}{\overset{\mathrm{rep}}{\mp}} \frac{1}{A_{\mathrm{free}}} \sqrt{\frac{8k_{\mathrm{cl}}k_{\mathrm{B}}T}{Q_{\mathrm{cl}}\omega_0} B_{\mathrm{ENBW}}}. \tag{66}$$

This equation serves as a reference for discussing the theoretical framework of AM-AFM.

In addition, for an infinite $Q_{\mathrm{cl}}$, where $B_{\mathrm{ENBW}}$ is limited by the cantilever bandwidth, by substituting Eq. (46), it can also be expressed in a form independent of $B_{\mathrm{ENBW}}$ as follows:

$$\lim_{Q_{\mathrm{cl}}\to\infty} k_{\mathrm{ts,min}}^{\mathrm{lim}} \approx \underset{\mathrm{att}}{\overset{\mathrm{rep}}{\mp}} \frac{1}{A_{\mathrm{free}}} \frac{\sqrt{k_{\mathrm{cl}}k_{\mathrm{B}}T}}{Q_{\mathrm{cl}}}. \tag{67}$$

At the very beginning of AM-AFM development, McClelland et al. derived the theoretical expression of $k_{\mathrm{ts,min}}$ for AM-AFM ($k_{\mathrm{ts,min}}^{\mathrm{AM}}$) when the sample, rather the cantilever, is excited at $f_0$, as follows [2]:

$$k_{\mathrm{ts,min}}^{\mathrm{AM,McClelland}} = \underset{\mathrm{att}}{\overset{\mathrm{rep}}{\mp}} \frac{1}{A_{\mathrm{sample}}} \sqrt{\frac{4k_{\mathrm{cl}}k_{\mathrm{B}}T}{Q_{\mathrm{cl}}\omega_0} \left(2B_{\mathrm{ENBW}}\right)}, \tag{68}$$

where $A_{\mathrm{sample}}$ denotes the oscillation amplitude of the sample. In this equation, unlike in the original literature, a factor of $\sqrt{2}$ is included to account for the two sidebands. If this factor is not included, the value is underestimated by a factor of 0.71 ($=\sqrt{1/2}$) compared to the dynamic-mode limit (Eq. (66)), as mentioned in Section 1. This adjustment renders the theoretical equations for AM-AFM and the dynamic-mode limit equivalent.

However, in the sample excitation method, although the increase in $A_{\mathrm{cl}}$ resulting from the sample oscillation is used as the feedback signal, $F_{\mathrm{ts}}$ simultaneously causes a decrease in $A_{\mathrm{cl}}$ during actual measurements. This expression does not account for this effect and therefore the equation inevitably overestimates the actual value. Furthermore, because of this effect, $A_{\mathrm{cl}}$ does not increase monotonically with the tip–sample distance, resulting in unstable imaging. For this reason, most current AM-AFM systems employ a cantilever-excitation scheme instead, and this theoretical expression is therefore not applicable to them.

In addition, Martin et al. derived an analytical solution for $k_{\text{ts,min}}^{\text{AM}}$ in the cantilever-excitation system [1]. In the infinity $Q_{\text{cl}}$ limit, $A_{\text{cl}}$ in Eq. (4) can be approximated by a Lorentzian function as follows:

$$\lim_{Q_{\text{cl}}\to\infty} A_{\text{cl}}(\tilde{\omega}_{\text{drive}}) \approx \frac{A_0}{\sqrt{4\left(1-\tilde{\omega}_{\text{drive}}\right)^2 + 1/Q_{\text{cl}}^2}}. \tag{69}$$

Since the force detection sensitivity in AM-AFM is maximized through excitation at the resonance slope, they derived it at the MaxSlope frequency, at which the amplitude slope reaches its maximum. Analytical approximations of the LMS and UMS frequencies in the infinity $Q_{\text{cl}}$ limit are obtained by differentiating Eq. (69) with respect to $f_{\text{drive}}$ and solving it as follows [1,59]:

$$\lim_{Q_{\text{cl}}\to\infty} \tilde{\omega}_{\text{LMS}} = \lim_{Q_{\text{cl}}\to\infty} \frac{f_{\text{LMS}}}{f_0} \approx 1 - \frac{1}{\sqrt{8}Q_{\text{cl}}}, \tag{70}$$

$$\lim_{Q_{\text{cl}}\to\infty} \tilde{\omega}_{\text{UMS}} = \lim_{Q_{\text{cl}}\to\infty} \frac{f_{\text{UMS}}}{f_0} \approx 1 + \frac{1}{\sqrt{8}Q_{\text{cl}}}. \tag{71}$$

Using these equations, the theoretical expression of $k_{\text{ts,min}}^{\text{AM}}$ was derived as follows [1]:

$$\begin{aligned}\lim_{Q_{\text{cl}}\to\infty} k_{\text{ts,min}}^{\text{AM,Martin}} &\overset{\text{rep}}{\underset{\text{att}}{=\mp}} \frac{1}{A_{\text{free}}}\sqrt{\frac{27k_{\text{cl}}k_{\text{B}}T}{Q_{\text{cl}}\omega_0}\left(2B_{\text{ENBW}}\right)} \\ &= \frac{3\sqrt{3}}{2}k_{\text{ts,min}}^{\text{lim}} \approx 2.60k_{\text{ts,min}}^{\text{lim}}.\end{aligned} \tag{72}$$

Unlike the original equation, we account for the factor of $\sqrt{2}$. If this coefficient is not included, the value is overestimated by a factor of 1.84 ($=\sqrt{27/4}$) compared to the dynamic-mode limit (Eq. (66)), as mentioned in Section 1. However, this equation does not produce exact MDF because it was derived under several assumptions, as detailed below. (1) Although the equation assumes excitation at the MaxSlope frequencies, it was obtained by substituting the noise expression at $f_0$. (2) MDF is estimated solely from the resonance frequency shift resulting from $F_{\text{ts}}$ ($\Delta f_{\text{ts}}$), assuming that $\Delta f_{\text{ts}} / \Delta A_{\text{ts}}$ remains constant even under the interaction. However, in reality, the $F_{\text{ts}}$ also changes the excitation efficiency, causing $\Delta f_{\mathrm{ts}} / \Delta A_{\mathrm{ts}}$ to vary (see Fig. 1 in Ref. [59]); therefore, this assumption does not hold. (3) As it is derived under the high $Q_{\mathrm{cl}}$ approximation, the equation is not applicable to liquid AM-AFM systems.

In more recent work, B. Voigtländer derived another analytical solution for $k_{\mathrm{ts,min}}^{\mathrm{AM}}$ in the cantilever-excitation system at the resonance slope, defined as follows:

$$\tilde{\omega}_{1/2} = 1 \underset{\mathrm{att}}{\overset{\mathrm{rep}}{\mp}} \frac{1}{2Q_{\mathrm{cl}}}. \tag{73}$$

This expression is equivalent to the MinForce frequencies (Eqs. (20) and (21)) in the limit of infinite $Q_{\mathrm{cl}}$. Similar to Martin's approach, by employing a Lorentzian approximation (Eq. (69)), the theoretical expression for $k_{\mathrm{ts,min}}^{\mathrm{AM}}$ was derived as follows [5]:

$$\begin{aligned} \lim_{Q_{\mathrm{cl}} \to \infty} k_{\mathrm{ts,min}}^{\mathrm{AM,Voigtländer}} &= \underset{\mathrm{att}}{\overset{\mathrm{rep}}{\mp}} \frac{1}{A_{\mathrm{free}}} \sqrt{\frac{8k_{\mathrm{cl}}k_{\mathrm{B}}T}{Q_{\mathrm{cl}}\omega_0}\left(2B_{\mathrm{ENBW}}\right)} \\ &= \sqrt{2}k_{\mathrm{ts,min}}^{\mathrm{lim}}. \end{aligned} \tag{74}$$

Unlike Martin et al., the author takes into account the $f_{\mathrm{drive}}$ dependence of the noise and the $\sqrt{2}$ factor. However, the derivation still assumes that $\Delta f_{\mathrm{ts}} / \Delta A_{\mathrm{ts}}$ remains constant. Furthermore, the author assumed that $\Delta f_{\mathrm{ts}} / \Delta A_{\mathrm{ts}}$ can be expressed as follows:

$$\frac{\Delta f_{\mathrm{ts}}}{\Delta A_{\mathrm{ts}}} \approx \frac{f_0}{Q_{\mathrm{cl}} A_{\mathrm{free}}}. \tag{75}$$

However, differentiating Eq. (69) rigorously and substituting Eq. (73) into it yield the equation with a $\sqrt{2}$ coefficient as follows:

$$\frac{\Delta f_{\mathrm{ts}}}{\Delta A_{\mathrm{ts}}} = \sqrt{2}\frac{f_0}{Q_{\mathrm{cl}} A_{\mathrm{free}}}. \tag{76}$$

Thus, this expression was also derived under several assumptions and does not appear to provide an exact solution.

## 9. Driving frequency dependent MDF in AM-AFM

As shown in the previous section, $k_{\mathrm{ts,min}}^{\mathrm{AM}}$ from the past studies does not appear to yield exact solutions due to being derived under several assumptions. Therefore, in this section, by integrating the insights gained in the preceding sections, we construct a theoretical framework that enables precise prediction of MDF in AM-AFM. Hereafter, MDF collectively refers to $k_{\mathrm{ts,min}}$ and $\langle F_{\mathrm{ts}}\rangle_{\mathrm{min}}$. Since MDF can be obtained by setting $\Delta A_{\mathrm{ts}} = N_{\mathrm{th}}^{\mathrm{AM}}(\tilde{\omega}_{\mathrm{drive}})$ [1,2], substituting the accurate noise equation (Eq. (44)), which rigorously accounts for the $f_{\mathrm{drive}}$ dependency, into linear approximations (Eq. (14)) yields

$$k_{\mathrm{ts,min}}^{\mathrm{AM,linear}} \overset{\mathrm{rep}}{\underset{\mathrm{att}}{=\mp}} \frac{1}{A_{\mathrm{free}}}\sqrt{\frac{4k_{\mathrm{cl}}k_{\mathrm{B}}T}{Q_{\mathrm{cl}}\omega_0}\left\{1+\left[\frac{\tilde{\omega}_{\mathrm{drive}}}{Q_{\mathrm{cl}}\left(1-\tilde{\omega}_{\mathrm{drive}}^2\right)}\right]^2\right\}\left(2B_{\mathrm{ENBW}}\right)}, \quad \text{for } \tilde{\omega}_{\mathrm{drive}} \neq 1, \tag{77}$$

Expressed in terms of the dynamic-mode limit (Eq. (66)), their difference is clearly highlighted as follows:

$$k_{\mathrm{ts,min}}^{\mathrm{AM,linear}} = \sqrt{1+\left[\frac{\tilde{\omega}_{\mathrm{drive}}}{Q_{\mathrm{cl}}\left(1-\tilde{\omega}_{\mathrm{drive}}^2\right)}\right]^2}\, k_{\mathrm{ts,min}}^{\mathrm{lim}}, \quad \text{for } \tilde{\omega}_{\mathrm{drive}} \neq 1, \tag{78}$$

Under the linear approximation, $\langle F_{\mathrm{ts}}\rangle_{\mathrm{min}}$ for AM-AFM ($\langle F_{\mathrm{ts}}\rangle_{\mathrm{min}}^{\mathrm{AM}}$) can also be obtained from Eq. (18) as follows:

$$\langle F_{\mathrm{ts}}\rangle_{\mathrm{min}}^{\mathrm{AM,linear}} \overset{\mathrm{rep}}{\underset{\mathrm{att}}{=\pm}} \sqrt{\frac{k_{\mathrm{cl}}k_{\mathrm{B}}T}{Q_{\mathrm{cl}}\omega_0}\left\{1+\left[\frac{\tilde{\omega}_{\mathrm{drive}}}{Q_{\mathrm{cl}}\left(1-\tilde{\omega}_{\mathrm{drive}}^2\right)}\right]^2\right\}\left(2B_{\mathrm{ENBW}}\right)}, \quad \text{for } \tilde{\omega}_{\mathrm{drive}} \neq 1. \tag{79}$$

These equations are valid for $\tilde{\omega}_{\mathrm{drive}} < 1$ and $\tilde{\omega}_{\mathrm{drive}} > 1$ in the repulsive and attractive regimes, respectively. As is evident, $k_{\mathrm{ts,min}}^{\mathrm{AM}}$ has the same form as $\langle F_{\mathrm{ts}}\rangle_{\mathrm{min}}^{\mathrm{AM}}$ with only different coefficients as

follows:

$$\langle F_{\mathrm{ts}}\rangle_{\min}^{\mathrm{AM,linear}} = -\frac{A_{\mathrm{free}}}{2} k_{\mathrm{ts,min}}^{\mathrm{AM,linear}} . \tag{80}$$

Thus, the $\langle F_{\mathrm{ts}}\rangle_{\min}$ in the dynamic-mode limit, $\langle F_{\mathrm{ts}}\rangle_{\min}^{\mathrm{lim}}$, can also be expressed as follows:

$$\langle F_{\mathrm{ts}}\rangle_{\min}^{\mathrm{lim}} = -\frac{A_{\mathrm{free}}}{2} k_{\mathrm{ts,min}}^{\mathrm{lim}} \overset{\mathrm{rep}}{\underset{\mathrm{att}}{=\pm}} \sqrt{\frac{2k_{\mathrm{cl}}k_{\mathrm{B}}T}{Q_{\mathrm{cl}}\omega_0} B_{\mathrm{ENBW}}} . \tag{81}$$

Using this equation, $\langle F_{\mathrm{ts}}\rangle_{\min}^{\mathrm{AM,linear}}$ is described as follows:

$$\langle F_{\mathrm{ts}}\rangle_{\min}^{\mathrm{AM,linear}} = \sqrt{1+\left[\frac{\tilde{\omega}_{\mathrm{drive}}}{Q_{\mathrm{cl}}\left(1-\tilde{\omega}_{\mathrm{drive}}^2\right)}\right]^2} \langle F_{\mathrm{ts}}\rangle_{\min}^{\mathrm{lim}}, \quad \text{for } \tilde{\omega}_{\mathrm{drive}} \neq 1. \tag{82}$$

Thus, when normalized by the dynamic-mode limit, $k_{\mathrm{ts,min}}^{\mathrm{AM,linear}}$ and $\langle F_{\mathrm{ts}}\rangle_{\min}^{\mathrm{AM,linear}}$ share the same expression.

In Fig. 7(a,b), we examine the $\tilde{\omega}_{\mathrm{drive}}$ dependence of normalized MDF, where the solid lines represent the linear solutions. In the repulsive regime, MDF decreases with decreasing $\tilde{\omega}_{\mathrm{drive}}$, whereas in the attractive regime, it decreases with increasing $\tilde{\omega}_{\mathrm{drive}}$. This change becomes steeper as $Q_{\mathrm{cl}}$ increases. This result indicates that, under ideal conditions, the lowest MDF in the repulsive regime is achieved at $\tilde{\omega}_{\mathrm{drive}}$ = 0, whereas in the attractive regime, it is achieved as $\tilde{\omega}_{\mathrm{drive}}$ approaches infinity. Therefore, by substituting $\tilde{\omega}_{\mathrm{drive}} = 0$ into Eqs. (78) and (82), the lowest MDF for AM-AFM is obtained as follows:

$$\begin{cases} k_{\mathrm{ts,min}}^{\mathrm{AM,linear}}(\tilde{\omega}_{\mathrm{drive}} = 0) = k_{\mathrm{ts,min}}^{\mathrm{lim}}, \\ \langle F_{\mathrm{ts}}(\tilde{\omega}_{\mathrm{drive}} = 0)\rangle_{\min}^{\mathrm{AM,linear}} = \langle F_{\mathrm{ts}}\rangle_{\min}^{\mathrm{lim}} . \end{cases} \tag{83}$$

This indicates that it is equivalent to the dynamic-mode limit, demonstrating consistency between the present study and previous research.

However, since the linear approximation is invalid near $f_0$, we also calculate the exact nonlinear solution obtained by coupling Eq. (8) or (16) with Eq. (44). Since the thermal noise is generally much smaller than $A_{\text{free}}$, the term $1/\tilde{A}$ outside the bracket in Eq. (8) can be approximated as 1 with only negligible error, as demonstrated later in Fig. 8. Although the expressions of $k_{\text{ts,min}}^{\text{AM}}$ and $\langle F_{\text{ts}} \rangle_{\text{min}}^{\text{AM}}$ differ slightly in the nonlinear case, this approximation makes them identical when normalized by the dynamic-mode limit, as in the linear case. Dashed lines in Fig. 7(a,b) show the nonlinear solution calculated assuming typical HS-AFM conditions: $k_{\text{cl}}$ = 0.1 N/m, $f_0$ = 1 MHz, $Q_{\text{cl}}$ = 1.5, $A_{\text{free}}$ = 3 $\text{nm}_{\text{p–0}}$, and $B_{\text{ENBW}}$ = 60 kHz. As shown, unlike the linear approximation, the nonlinear solution does not diverge at $f_0$ but instead converges to a value several times the value of the dynamic-mode limit.

These results indicate that the MDF at $f_0$ significantly reflects the discrepancy between the linear approximation and the nonlinear solution. Therefore, we consider that it serves as an indicator of the agreement between the two, and we derive an analytical expression for the MDF at $f_0$. This expression is obtained by substituting the noise expression at $f_0$ (Eq. (33)) into Eqs. (10) and (17), as follows:

$$\begin{cases} k_{\text{ts,min}}^{\text{AM}}\left(\tilde{\omega}_{\text{drive}}=1\right) = \underset{\text{att}}{\overset{\text{rep}}{\mp}} \sqrt{\dfrac{2k_{\text{cl}}}{Q_{\text{cl}}}\left|k_{\text{ts,min}}^{\text{lim}}\right|}, \\ \left\langle F_{\text{ts}}\left(\tilde{\omega}_{\text{drive}}=1\right)\right\rangle_{\text{min}}^{\text{AM}} = \underset{\text{att}}{\overset{\text{rep}}{\pm}} \sqrt{\dfrac{k_{\text{cl}}A_{\text{free}}}{Q_{\text{cl}}}\left|\left\langle F_{\text{ts}}\right\rangle_{\text{min}}^{\text{lim}}\right|}. \end{cases} \tag{84}$$

Due to the square root dependence, the result deviates from a linear relationship with the dynamic-mode limit and varies with multiple parameters. To examine the parameter dependence, we derive the ratio to the dynamic-mode limit as follows:

$$\frac{k_{\text{ts,min}}^{\text{AM}}(\tilde{\omega}_{\text{drive}}=1)}{k_{\text{ts,min}}^{\text{lim}}} = \frac{\left\langle F_{\text{ts}}\left(\tilde{\omega}_{\text{drive}}=1\right)\right\rangle_{\text{min}}^{\text{AM}}}{\left\langle F_{\text{ts}}\right\rangle_{\text{min}}^{\text{lim}}} = \sqrt{A_{\text{free}}\sqrt{\frac{k_{\text{cl}}\omega_0}{2Q_{\text{cl}}k_{\text{B}}TB_{\text{ENBW}}}}}, \tag{85}$$

As this value increases, the nonlinear solution approaches the linear approximation, whereas as it decreases, it approaches the dynamic-mode limit. In this expression, $\omega_0/2Q_{\text{cl}}$ and $B_{\text{ENBW}}$ are generally correlated. In particular, at high $Q_{\text{cl}}$, the cantilever bandwidth limits the system, and substituting Eq. (46) yields a more explicit expression as follows:

$$\lim_{Q_{\rm cl}\to\infty}\frac{k_{\rm ts,min}^{\rm AM}(\tilde{\omega}_{\rm drive}=1)}{k_{\rm ts,min}^{\rm lim}}=\lim_{Q_{\rm cl}\to\infty}\frac{\left\langle F_{\rm ts}\left(\tilde{\omega}_{\rm drive}=1\right)\right\rangle_{\rm min}^{\rm AM}}{\left\langle F_{\rm ts}\right\rangle_{\rm min}^{\rm lim}}=\sqrt{2A_{\rm free}\sqrt{\frac{k_{\rm cl}}{k_{\rm B}T}}}, \tag{86}$$

Consequently, we found that the consistency between the linear and nonlinear solutions can be represented as a function of $A_{\rm free}\sqrt{k_{\rm cl}}$ . Accordingly, we calculate the $\tilde{\omega}_{\rm drive}$ dependence of MDF at $Q_{\rm cl}$ = 1.5 for $A_{\rm free}\sqrt{k_{\rm cl}}$ in the range of 1 to $9\,{\rm nm}\sqrt{\rm N/m}$, which represents typical conditions for liquid and ambient experiments, as described in Section 4. In Fig. 7(c), the calculated results clearly indicate that as $A_{\rm free}\sqrt{k_{\rm cl}}$ increases (corresponding to an increase in the SNR), the nonlinear solution asymptotically approaches the linear approximation. Conversely, this implies that decreasing $A_{\rm free}\sqrt{k_{\rm cl}}$ brings the system closer to the dynamic-mode limit.

## 10. Experimental validation of frequency dependent MDF

This section presents experimental validation of the theoretical $f_{\mathrm{drive}}$ dependence of the MDF. Before evaluating the MDF, we analyze the $f_{\mathrm{drive}}$-dependence of the normalized noise-equivalent amplitude $\tilde{A}_{\mathrm{noise}}$, which is defined as follows:

$$\tilde{A}_{\mathrm{noise}} \equiv \frac{N_{\mathrm{th}}^{\mathrm{AM}}(\tilde{\omega}_{\mathrm{drive}})}{A_{\mathrm{free}}}. \tag{87}$$

Substituting Eq. (44) gives the SHO expression as follows:

$$\tilde{A}_{\mathrm{noise}} = \frac{1}{A_{\mathrm{free}}}\sqrt{\frac{4k_{\mathrm{B}}TQ_{\mathrm{cl}}}{k_{\mathrm{cl}}\omega_0}\frac{\left(2B_{\mathrm{ENBW}}\right)}{\left[Q_{\mathrm{cl}}\left(1-\tilde{\omega}_{\mathrm{drive}}^2\right)\right]^2+\tilde{\omega}_{\mathrm{drive}}^2}}. \tag{88}$$

Figure 7(d) shows the experimental results obtained with the same experimental setup described in Section 7, with a cantilever (BLAC10DS-A2, Olympus) having $f_0$ of 660 kHz, $Q_{\mathrm{cl}}$ of 1.7, and $k_{\mathrm{cl}}$ of 0.13 N/m. As shown, $\tilde{A}_{\mathrm{noise}}$ corresponded to 2–3% of $A_{\mathrm{free}}$. Reflecting the original deflection noise characteristics, it exhibited a maximum at $f_{\mathrm{peak}}$ and a slightly broader peak than the SHO-fitted result (blue dashed curve).

Figure 7(e) shows the $f_{\mathrm{drive}}$-dependence of experimental $\langle F_{\mathrm{ts}}\rangle_{\mathrm{min}}^{\mathrm{AM}}$, compared with the analytical calculations. For the experimental results, two curves are presented: one calculated from the experimentally obtained noise (Fig. 7(d)), and the other derived from a SHO fit to this noise spectrum. They overlap with each other with at most an 11% deviation. These experimental results clearly demonstrated that $\langle F_{\mathrm{ts}}\rangle_{\mathrm{min}}^{\mathrm{AM}}$ is strongly positively correlated with $f_{\mathrm{drive}}$, as theoretically predicted. However, the data revealed that, unlike the theoretical prediction, $\langle F_{\mathrm{ts}}\rangle_{\mathrm{min}}^{\mathrm{AM}}$ exhibited a slight local minimum near the resonance slope. Compared with the theoretical dynamic-mode limit (green dashed line, Eq. (81)), this minimum value remains slightly larger, confirming that Eq. (81)

indeed represents the theoretical lower limit. Compared with the Martin approximation (purple dashed line, Eq. (72)), it takes approximately half the value, confirming that the Martin approximation significantly overestimates the MDF.

In contrast, the theoretical curve shows a monotonic increase with $f_{\mathrm{drive}}$ but agrees with the experimental results near the resonance slope. In the frequency range where $\tilde{\omega}_{\mathrm{drive}} < 1$, the experimental and theoretical results showed reasonable agreement. Although the slight deviation was observed, it is considered to be due to the use of acoustic excitation in this experiment, whereby the vibration amplitude of the cantilever base becomes comparable to that of the cantilever tip, especially at low frequencies [84]. However, for $\tilde{\omega}_{\mathrm{drive}} > 1$, a significant discrepancy was observed, which is considered to result from energy dissipation, such as adhesion. These discrepancies will be discussed in more detail elsewhere.

## 11. MDF vs $Q_{cl}$ under linear approximation

Since AM-AFM operates in air and liquid environments, $Q_{cl}$ varies widely from about 1 to 300 depending on the viscosity of the medium. Therefore, in this section, we analyze the $Q_{cl}$-dependence of MDF to identify the optimum condition where MDF is minimized. As discussed in our previous study [59], from the perspective of non-destructive observation of fragile molecules, excitation at the resonance slope is important. In this frequency range, the linear approximations for $k_{ts}$ and $\langle F_{ts} \rangle$ are valid. Thus, by substituting Eqs. (20) and (21) into Eq. (78), simplified analytical expressions of $k_{ts,min}^{AM}$ for $f_{LMF}$ in the repulsive regime and for $f_{UMF}$ in the attractive regime can be obtained as follows:

$$k_{ts,min}^{AM,LMF} = \sqrt{2 - \frac{1}{Q_{cl}}} k_{ts,min}^{lim}, \quad \text{for} \langle F_{ts} \rangle > 0, Q_{cl} \geq 1, \tag{89}$$

$$k_{ts,min}^{AM,UMF} = \sqrt{2 + \frac{1}{Q_{cl}}} k_{ts,min}^{lim}, \quad \text{for} \langle F_{ts} \rangle < 0. \tag{90}$$

Similarly, substituting Eqs. (20) and (21) into Eq. (82) gives simplified analytical equations of $\langle F_{ts} \rangle_{min}^{AM}$ for $f_{LMF}$ and $f_{UMF}$ as follows:

$$\langle F_{ts} \rangle_{min}^{AM,LMF} = \sqrt{2 - \frac{1}{Q_{cl}}} \langle F_{ts} \rangle_{min}^{lim}, \quad \text{for} \langle F_{ts} \rangle > 0, Q_{cl} \geq 1, \tag{91}$$

$$\langle F_{ts} \rangle_{min}^{AM,UMF} = \sqrt{2 + \frac{1}{Q_{cl}}} \langle F_{ts} \rangle_{min}^{lim}, \quad \text{for} \langle F_{ts} \rangle < 0. \tag{92}$$

In Fig. 8(a), we examine the $Q_{cl}$-dependence of normalized MDF. As shown, under the infinite $Q_{cl}$ condition, both the coefficients for LMF and UMF asymptotically approach $\sqrt{2}$ as follows:

$$
\underbrace{\frac{\lim_{Q_{\mathrm{cl}}\to\infty} k_{\mathrm{ts,min}}^{\mathrm{AM,LMF}}}{k_{\mathrm{ts,min}}^{\mathrm{lim}}} = \frac{\lim_{Q_{\mathrm{cl}}\to\infty} \langle F_{\mathrm{ts}} \rangle_{\mathrm{min}}^{\mathrm{AM,LMF}}}{\langle F_{\mathrm{ts}} \rangle_{\mathrm{min}}^{\mathrm{lim}}}}_{\text{repulsive regime}} = \underbrace{\frac{\lim_{Q_{\mathrm{cl}}\to\infty} k_{\mathrm{ts,min}}^{\mathrm{AM,UMF}}}{k_{\mathrm{ts,min}}^{\mathrm{lim}}} = \frac{\lim_{Q_{\mathrm{cl}}\to\infty} \langle F_{\mathrm{ts}} \rangle_{\mathrm{min}}^{\mathrm{AM,UMF}}}{\langle F_{\mathrm{ts}} \rangle_{\mathrm{min}}^{\mathrm{lim}}}}_{\text{attractive regime}} = \sqrt{2}. \tag{93}
$$

In contrast, as $Q_{\mathrm{cl}}$ decreases, the coefficient for LMF decreases toward 1, corresponding to the dynamic-mode limit, whereas that for UMF increases toward 2.

These results clearly indicate that the MDF of AM-AFM takes values greater than the dynamic-mode limit across the entire $Q_{\mathrm{cl}}$ range, especially at high $Q_{\mathrm{cl}}$. This is because, whereas FM-AFM can detect purely conservative forces, AM-AFM measures a signal mixed with dissipation, which in turn reduces its sensitivity to conservative interactions.

In contrast, this result indicates that, under liquid conditions with low $Q_{\mathrm{cl}}$, a minimum MDF approaching the theoretical limit can be achieved. Using typical HS-AFM conditions: $k_{\mathrm{cl}} = 0.1$ N/m, $f_0 = 1$ MHz, $Q_{\mathrm{cl}} = 1.5$, and $B_{\mathrm{ENBW}} = 60$ kHz, the coefficient for LMF is estimated to be 1.15, resulting in $\langle F_{\mathrm{ts}} \rangle_{\mathrm{min}}^{\mathrm{AM,LMF}}$ of 2.6 pN. Since biomolecular imaging is typically conducted at around 10 pN, this value is highly realistic.

Furthermore, we found that the Voigtländer approximation (Eq. (74)) also yields a coefficient of $\sqrt{2}$, which coincides with that of our exact solution in the infinite $Q_{\mathrm{cl}}$ limit. This agreement is considered to be coincidental, as the neglected dependence on excitation efficiency and the factor of $\sqrt{2}$ effectively cancel each other out.

Meanwhile, although the theoretical limit of MDF in AM-AFM can be obtained by excitation at the MinForce frequencies, excitation at the MaxSlope frequencies is also commonly used in actual measurements. Therefore, we also examine the $Q_{\mathrm{cl}}$ dependence of MDF for MaxSlope. Since an analytical expression for the MaxSlope frequency cannot be obtained, the previously derived Laurent polynomial approximation is used [59]. In Fig. 8(a), the MaxSlope results exhibit a dependency on $Q_{\mathrm{cl}}$ similar to that of MinForce but are slightly higher. Under the infinite $Q_{\mathrm{cl}}$ condition, both the coefficients for LMF and UMF asymptotically approach $\sqrt{3}$ as follows:

$$\underbrace{\frac{\lim_{Q_{\text{cl}}\to\infty} k_{\text{ts,min}}^{\text{AM,LMS}}}{k_{\text{ts,min}}^{\text{lim}}} = \frac{\lim_{Q_{\text{cl}}\to\infty} \left\langle F_{\text{ts}} \right\rangle_{\text{min}}^{\text{AM,LMS}}}{\left\langle F_{\text{ts}} \right\rangle_{\text{min}}^{\text{lim}}}}_{\text{repulsive regime}} = \underbrace{\frac{\lim_{Q_{\text{cl}}\to\infty} k_{\text{ts,min}}^{\text{AM,UMS}}}{k_{\text{ts,min}}^{\text{lim}}} = \frac{\lim_{Q_{\text{cl}}\to\infty} \left\langle F_{\text{ts}} \right\rangle_{\text{min}}^{\text{AM,UMS}}}{\left\langle F_{\text{ts}} \right\rangle_{\text{min}}^{\text{lim}}}}_{\text{attractive regime}} = \sqrt{3}. \tag{94}$$

As $Q_{\text{cl}}$ decreases, the coefficient for LMS decreases toward 1, whereas the coefficient for UMS increases sharply by several times. As previously demonstrated, the force detection sensitivity at the MinForce frequencies is higher than at the MaxSlope frequencies by only 3% [59]; in contrast, by comparing Eqs. (93) and (94), we found that the difference is increased to 22% (from $\sqrt{2}$ to $\sqrt{3}$) in terms of MDF.

Furthermore, comparison with the Martin approximation (Eq. (72)), which was derived for the MaxSlope, shows that it overestimates the value by approximately a factor of 2 for all $Q_{\text{cl}}$ values. In particular, when compared with the LMS and UMS results in the limit of infinite $Q_{\text{cl}}$, it can be seen that the Martin approximation overestimates by a factor of 3/2 since it does not account for the excitation-efficiency contribution.

## 12. Comparison between dynamic and static modes

In this section, we compare MDF between static-mode and dynamic-mode AFM. The noise equation for static-mode AFM is derived by integrating the Brownian noise density in the low-frequency range, as follows:

$$N_{\mathrm{th}}^{\mathrm{static}} = \sqrt{\frac{4k_{\mathrm{B}}T}{k_{\mathrm{cl}}Q_{\mathrm{cl}}\omega_0}B_{\mathrm{ENBW}}}\,. \tag{95}$$

Coupling this equation with Hooke's law yields MDF for the static mode as follows [52]:

$$F_{\mathrm{min}}^{\mathrm{static}} = \sqrt{\frac{4k_{\mathrm{cl}}k_{\mathrm{B}}T}{Q_{\mathrm{cl}}\omega_0}B_{\mathrm{ENBW}}}\,. \tag{96}$$

For this derivation, the dynamic-to-static ratio of spring constant is assumed to 1. Meanwhile, to compare with the theoretical limit of $\langle F_{\mathrm{ts}}\rangle_{\mathrm{min}}^{\mathrm{AM}}$, substituting $Q_{\mathrm{cl}} = 1$ into Eq. (91) yields the equation as follows:

$$\langle F_{\mathrm{ts}}\rangle_{\mathrm{min}}^{\mathrm{lim}} = \langle F_{\mathrm{ts}}(Q_{\mathrm{cl}}=1)\rangle_{\mathrm{min}}^{\mathrm{AM,LMF}} = \sqrt{\frac{k_{\mathrm{cl}}k_{\mathrm{B}}T}{\omega_0 Q_{\mathrm{cl}}}(2B_{\mathrm{ENBW}})} \tag{97}$$

Comparison between Eqs. (96) and (97) indicates that the MDF of AM-AFM is $\sqrt{2}$ times smaller than static-mode AFM, meaning that estimating the cantilever's mean deflection from $\Delta A_{\mathrm{ts}}$ yields higher SNR than determining it directly. Although this may seem counterintuitive, it can be interpreted as follows. At $Q_{\mathrm{cl}} = 1$ or off-resonance, when the cantilever oscillates sinusoidally, only the lower half of the oscillation is affected by $F_{\mathrm{ts}}$, while the upper half remains unchanged. As a result, the mean deflection changes by only half of $\Delta A_{\mathrm{ts}}$. Meanwhile, the noise increases by $\sqrt{2}$ due to contributions from both sidebands, resulting in an overall $\sqrt{2}$ reduction in MDF. In brief, this discrepancy occurs because AM-AFM measures $\langle F_{\mathrm{ts}}\rangle$ rather than the instantaneous $F_{\mathrm{ts}}$.

## 13. MDF vs $Q_{cl}$ in nonlinear solution

In this section, we analyze the $Q_{cl}$-dependence of MDF for characteristic frequencies using the nonlinear solution by coupling Eqs. (8) and (16) with Eq. (44). The calculation conditions are identical to those used in the previous section. As described in Section 5, $B_{ENBW}$ are calculated using Eq. (50) by setting $\delta_{amp,z}$ = 29.5 µs so that $B_{ENBW}$ = 60 kHz at $Q_{cl}$ = 1.5. Although the latency of the amplitude detector slightly varies depending on $f_{drive}$, this effect is neglected.

Comparisons are performed for $A_{free}\sqrt{k_{cl}}$ values ranging from 1 to 9 $\mathrm{nm}\sqrt{\mathrm{N/m}}$. As described in Sections 4 and 9, since $A_{free}\sqrt{k_{cl}}$ = 1 represents typical liquid conditions, the calculation results are shown with an upper limit of $Q_{cl}$ = 10.

Fig. 8(b,c) shows results at $f_{LMF}$ and $f_{UMF}$, where the nonlinear solutions of both $k_{ts,min}^{AM}$ and $\langle F_{ts}\rangle_{min}^{AM}$ closely match the linear approximation across all values of $A_{free}\sqrt{k_{cl}}$ with an error of at most approximately 2%. Notably, as $A_{free}\sqrt{k_{cl}}$ increases, the results further converge toward the linear approximation, which confirms that the linear approximation can be effectively used at $f_{LMF}$ and $f_{UMF}$ for arbitrary $A_{free}\sqrt{k_{cl}}$.

Subsequently, in Fig. 8(d,e), the nonlinear solutions are also employed to compute MDF at $f_0$ and $f_{peak}$, where the linear approximation is no longer valid. Even in these cases, the values of $k_{ts,min}^{AM}$ and $\langle F_{ts}\rangle_{min}^{AM}$ closely match across all $Q_{cl}$ within approximately 5% at most. However, in contrast to LMF and UMF, however, the curves shows substantial variation depending on $A_{free}\sqrt{k_{cl}}$.

At $f_0$ [Fig. 8(d)], similar to UMF, MDF decreases with increasing $Q_{cl}$ for all values of $A_{free}\sqrt{k_{cl}}$. For $Q_{cl}$ values above ~32, the decrease saturates and it asymptotically approaches the $f_{cl}$-limited values given by Eq. (86) (9.67 for 3 $\mathrm{nm}\sqrt{\mathrm{N/m}}$ and 16.8 for 9 $\mathrm{nm}\sqrt{\mathrm{N/m}}$). At $A_{\mathrm{free}}\sqrt{k_{\mathrm{cl}}}$ = 1 $\mathrm{nm}\sqrt{\mathrm{N/m}}$, corresponding to typical liquid conditions, the MDF is approximately 10 times greater than the dynamic-mode limit. As $Q_{\mathrm{cl}}$ increases, a larger $A_{\mathrm{free}}\sqrt{k_{\mathrm{cl}}}$ is required, and consequently, the MDF remains more than 10 times larger than the dynamic-mode limit. These results suggest that, for a given $Q_{\mathrm{cl}}$, the MDF at $f_0$ can be reduced by minimizing $A_{\mathrm{free}}$ and $k_{\mathrm{cl}}$, as long as jump-to-contact is avoided. Note that equivalent results can also be readily calculated using Eq. (85).

At $f_{\mathrm{peak}}$ [Fig. 8(e)], unlike $f_0$, the MDF increases with increasing $Q_{\mathrm{cl}}$. However, at high $Q_{\mathrm{cl}}$, the increase saturates and it asymptotically approaches the value given by Eq. (86), similar to $f_0$. This equivalence at high $Q_{\mathrm{cl}}$ arises because, as seen from Eq. (11), $f_{\mathrm{peak}}$ approaches $f_0$ as $Q_{\mathrm{cl}}$ increases. In contrast, at low $Q_{\mathrm{cl}}$ conditions, the MDF remains at most 5 times the dynamic-mode limit regardless of $A_{\mathrm{free}}\sqrt{k_{\mathrm{cl}}}$, indicating that relatively high force resolution can be achieved in liquid environments. Notably, at $Q_{\mathrm{cl}}$ = 1.5, which is typical for HS-AFM, MDF is suppressed to approximately 2.5 times, comparable to the Martin approximation.

This analytical solution at $f_{\mathrm{peak}}$ within the low $Q_{\mathrm{cl}}$ range can be derived by coupling Eqs. (14) and (18) with Eqs. (11) and (44) as follows.

$$\begin{cases} \lim\limits_{Q_{\mathrm{cl}}\to 0} k_{\mathrm{ts,min}}^{\mathrm{AM}}\left(\tilde{\omega}_{\mathrm{drive}}=\tilde{\omega}_{\mathrm{peak}}\right)=\sqrt{4Q_{\mathrm{cl}}^2-1}\,k_{\mathrm{ts,min}}^{\mathrm{lim}}, \\ \lim\limits_{Q_{\mathrm{cl}}\to 0}\left\langle F_{\mathrm{ts}}\left(\tilde{\omega}_{\mathrm{drive}}=\tilde{\omega}_{\mathrm{peak}}\right)\right\rangle_{\mathrm{min}}^{\mathrm{AM}}=\sqrt{4Q_{\mathrm{cl}}^2-1}\left\langle F_{\mathrm{ts}}\right\rangle_{\mathrm{min}}^{\mathrm{lim}}. \end{cases} \tag{98}$$

These equations are only valid for $\langle F_{\mathrm{ts}}\rangle$ > 0. As shown by the gray dashed line in Fig. 8(e), this expression fits well in the region where $Q_{\mathrm{cl}} < 2$.

As shown in Fig. 4, the $Q_{\mathrm{cl}}$-dependence of $B_{\mathrm{ENBW}}$ varies depending on the system configuration. Therefore, we also investigate the effect of varying $\delta_{\mathrm{amp,z}}$. Calculation conditions are $A_{\mathrm{free}}\sqrt{k_{\mathrm{cl}}}$ = 3 $\mathrm{nm}\sqrt{\mathrm{N/m}}$ with $\delta_{\mathrm{amp,z}}$ ranging from 29.5 to 5000 μs. In Fig. 8(f), the results at $f_0$ and $f_{\mathrm{peak}}$ are shown by dashed and solid lines, respectively. As shown, increasing $\delta_{\mathrm{amp,z}}$ leads to an increase in MDF, particularly at low $Q_{cl}$ values. At higher $Q_{cl}$ values, the curves for $f_0$ and $f_{peak}$ asymptotically approach each other. As also shown in Fig. 4, increasing $\delta_{amp,z}$ results in a reduction in $B_{ENBW}$, which would typically be expected to reduce MDF as well. However, according to Eq. (85), the explicit value of MDF at $f_0$ is proportional to the 1/4 power of $B_{ENBW}$. Conversely, as shown in Eq. (86), the normalized MDF at $f_0$ is inversely proportional to the 1/4 power of $B_{ENBW}$. This result implies that, when operating at $f_0$ or $f_{peak}$, reducing $B_{ENBW}$ leads to only a limited improvement in MDF.

To summarize, under all $f_{drive}$ and $Q_{cl}$ conditions, MDF in AM-AFM was found to exceed the dynamic-mode limit, particularly beyond $f_{peak}$. This result can be interpreted as follows. The $I_{even}$ and $I_{odd}$ components shown in Eq. (1) are orthogonal in the complex plane; in principle, FM-AFM can detect them separately as frequency shift and energy dissipation, respectively [85]. Similarly, although the amplitude and phase are also orthogonal in polar coordinates, their detection vectors are not aligned with either $I_{even}$ or $I_{odd}$ components except for $f_0$. Since MDF theory considers only $I_{even}$, AM-AFM inherently yields larger MDF values than FM-AFM.

However, their sensitivities to $I_{even}$ and $I_{odd}$ depend on $f_{drive}$. At $f_0$, the amplitude sensitivity to $I_{even}$ is minimized while that to $I_{odd}$ is maximized. Conversely, off-resonance, the amplitude sensitivity to $I_{even}$ is maximized while that to $I_{odd}$ is minimized. This behavior explains the observed $f_{drive}$ dependence of MDF. Moreover, Since the theoretical MDF does not consider $I_{odd}$, the actual MDF in systems with high energy dissipation is expected to be somewhat lower. Since phase is orthogonal to amplitude, using both amplitude and phase to reconstruct the force [3,12,13,60] may achieve MDF values comparable to the dynamic-mode limit.

## 14. Implications for instrumentation improvement

Finally, as a key application of MDF, we discuss instrumentation improvements, focusing on differences between AM-AFM and other dynamic modes. Because the unnormalized MDF (Eq.(81)) value is inversely proportional to $\sqrt{Q_{\mathrm{cl}}}$ , in Fig. 9, we examine the dependence of explicit MDF ($\langle F_{\mathrm{ts}} \rangle_{\mathrm{min}}$) values on $Q_{\mathrm{cl}}$ for the dynamic-mode limit and AM-AFM at the resonance slope frequencies (LMF and LMS). All curves indicate that the MDF decreases inversely with the $Q_{\mathrm{cl}}$. In the high-$Q_{\mathrm{cl}}$ regime, the $Q_{\mathrm{cl}}$-dependence is the same except for differences in the proportionality coefficients; thus instrumentation requirements are essentially the same as for the other modes. In contrast, at low $Q_{\mathrm{cl}}$, clear differences appear. For the other modes, the MDF increases sharply below $Q_{\mathrm{cl}} \approx 2$, whereas for AM-AFM it tends to saturate, indicating reduced sensitivity to $Q_{\mathrm{cl}}$ in this range. Although AM-AFM benefits from higher response speed at lower $Q_{\mathrm{cl}}$, this typically increases the MDF, creating a trade-off. However, the present result suggests that this trade-off is mitigated under low-$Q_{\mathrm{cl}}$ conditions. Therefore, as demonstrated in recent HS-AFM developments [28], minimizing $Q_{\mathrm{cl}}$ and $k_{\mathrm{cl}}$ within stable imaging limits, while increasing $f_0$ and reducing the latency of the amplitude detector [77], leads to both lower MDF and higher scan speeds.

## 15. Conclusions

In this study, we derived an exact solution for the MDF in AM-AFM, resolving a long-standing uncertainty. While previous studies reported both smaller (better) and larger (worse) MDF values for AM-AFM relative to other dynamic-mode AFM, our analysis conclusively shows that AM-AFM performs worse under all conditions. Although at $f_0$, the MDF can be ten times greater than the dynamic-mode limit, it decreases toward this limit as $f_{\mathrm{drive}}$ decreases. When the cantilever is driven at the resonance slope, this ratio falls to 1.41 at high $Q_{\mathrm{cl}}$ and further decreases toward 1 as $Q_{\mathrm{cl}}$ decreases. At high $Q_{\mathrm{cl}}$ values, the ratio remains essentially constant; however, below $Q_{\mathrm{cl}} = 2$, the dependence on $Q_{\mathrm{cl}}$ weakens, resulting in behavior that is slightly distinct from other AFM modes. The theory constructed here provides a reliable benchmark for evaluating the absolute performance of instruments and offers guidelines for improving instrumentation.

# Figures

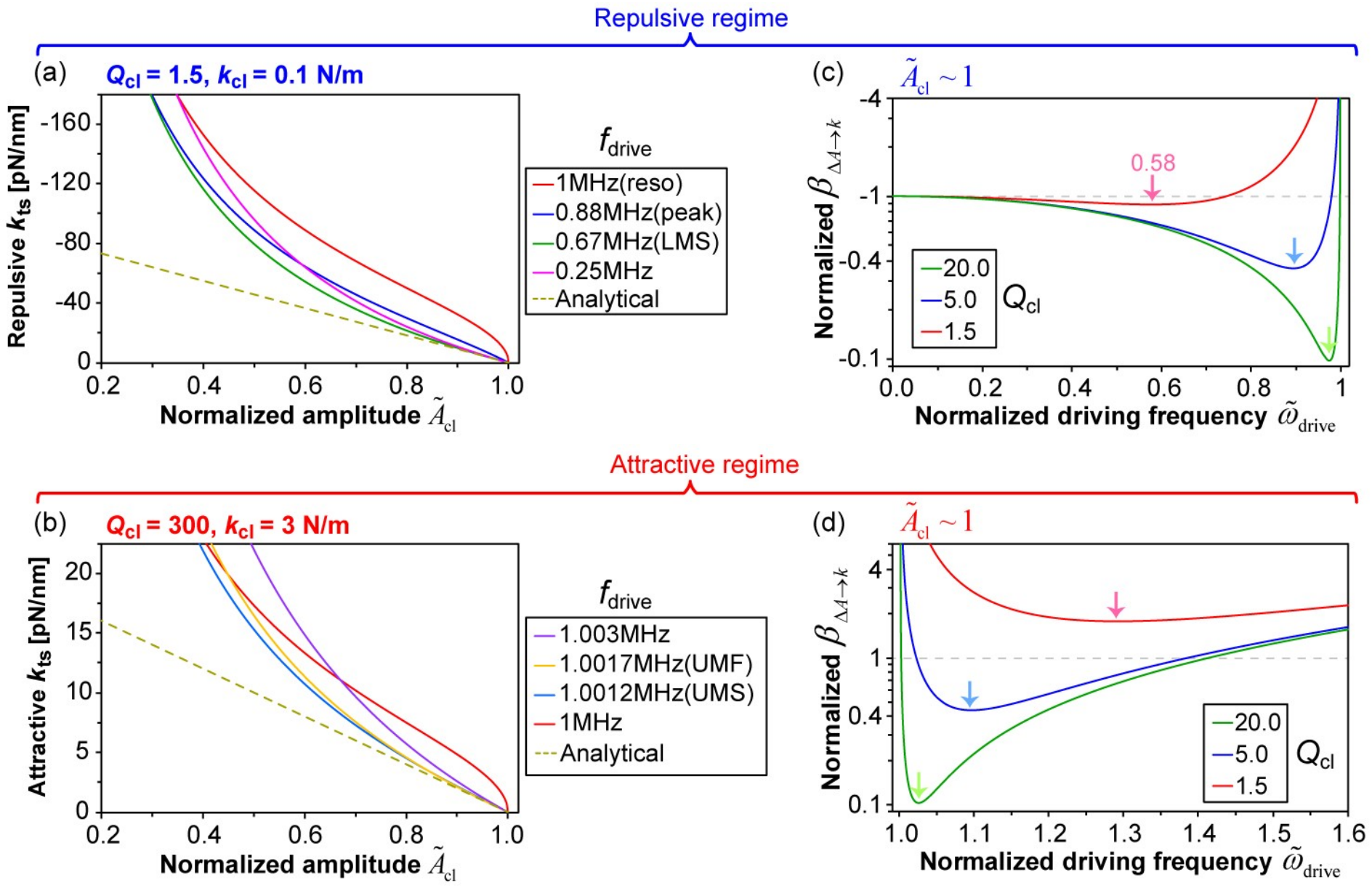

**FIG. 1.** (**a,b**) Theoretical normalized amplitude dependence of the force gradient at different $f_{\mathrm{drive}}$ in the repulsive (a) and attractive (b) regimes. In the legend, "reso", "peak", "LMS", "UMS", and "UMF" denote the resonance, peak, lower MaxSlope, upper MaxSlope, and upper MinForce frequencies, respectively. Orange dashed lines represent the linear approximation at LMS and UMF in panels (a) and (b), respectively. (**c,d**) $f_{\mathrm{drive}}$ dependence of $\beta_{\Delta A \to F'}$ ($\tilde{A}_{\mathrm{cl}} = 1$) normalized by those at $f_{\mathrm{drive}}$ = 0 Hz for various $Q_{\mathrm{cl}}$ in the repulsive (c) and attractive (d) regimes.

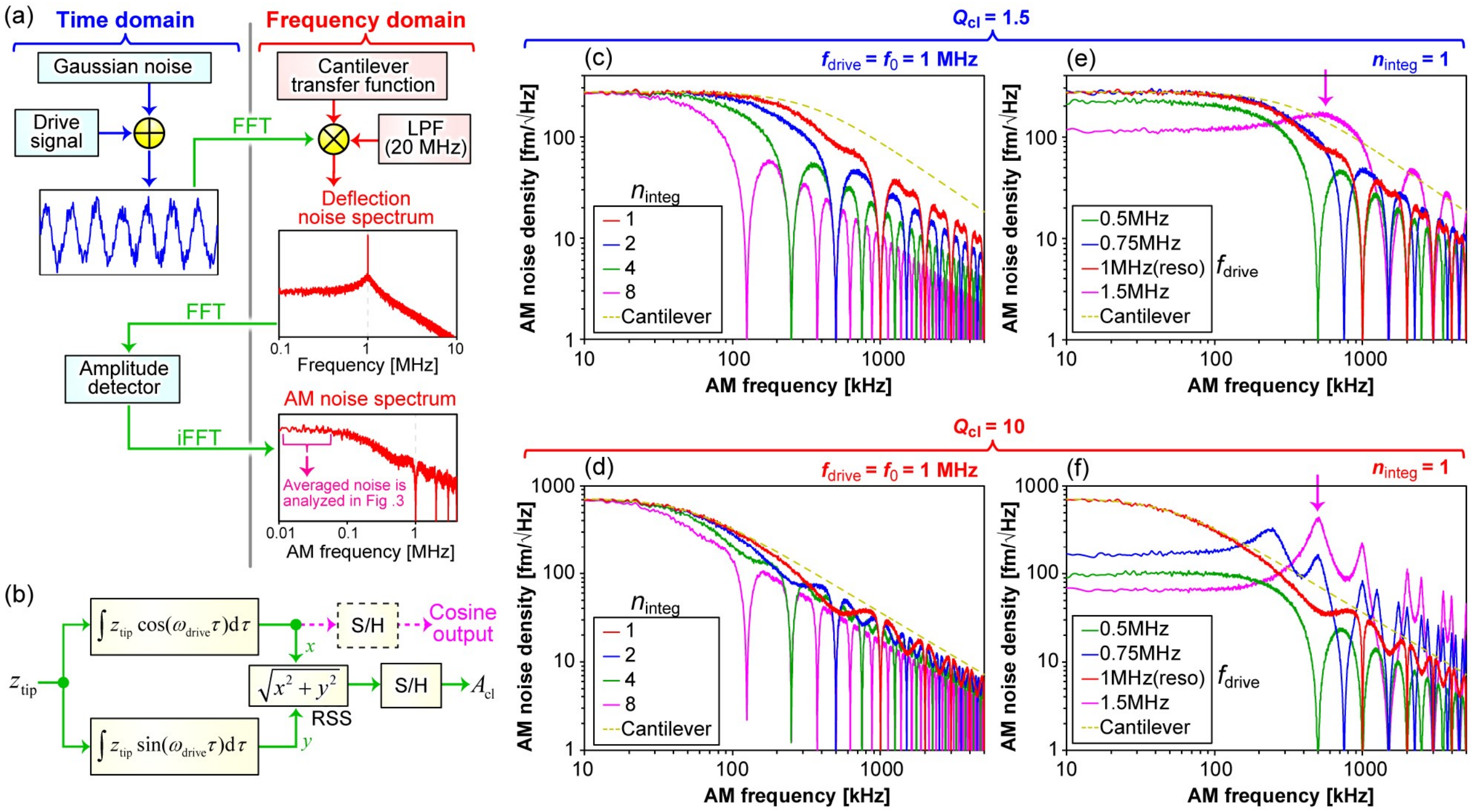

**FIG. 2.** (**a,b**) Block diagrams for simulating AM noise spectra in AM-AFM (a) and for the FAB amplitude detector (b). (**c–f**) Simulated AM noise density spectra using FAB, calculated for different $n_{\text{integ}}$ values (c,d) and different $f_{\text{drive}}$ values (e,f), for $Q_{\text{cl}}$ = 1.5 (c,e) and $Q_{\text{cl}}$ = 10 (d,f). The arrows indicate the gain noise peak.

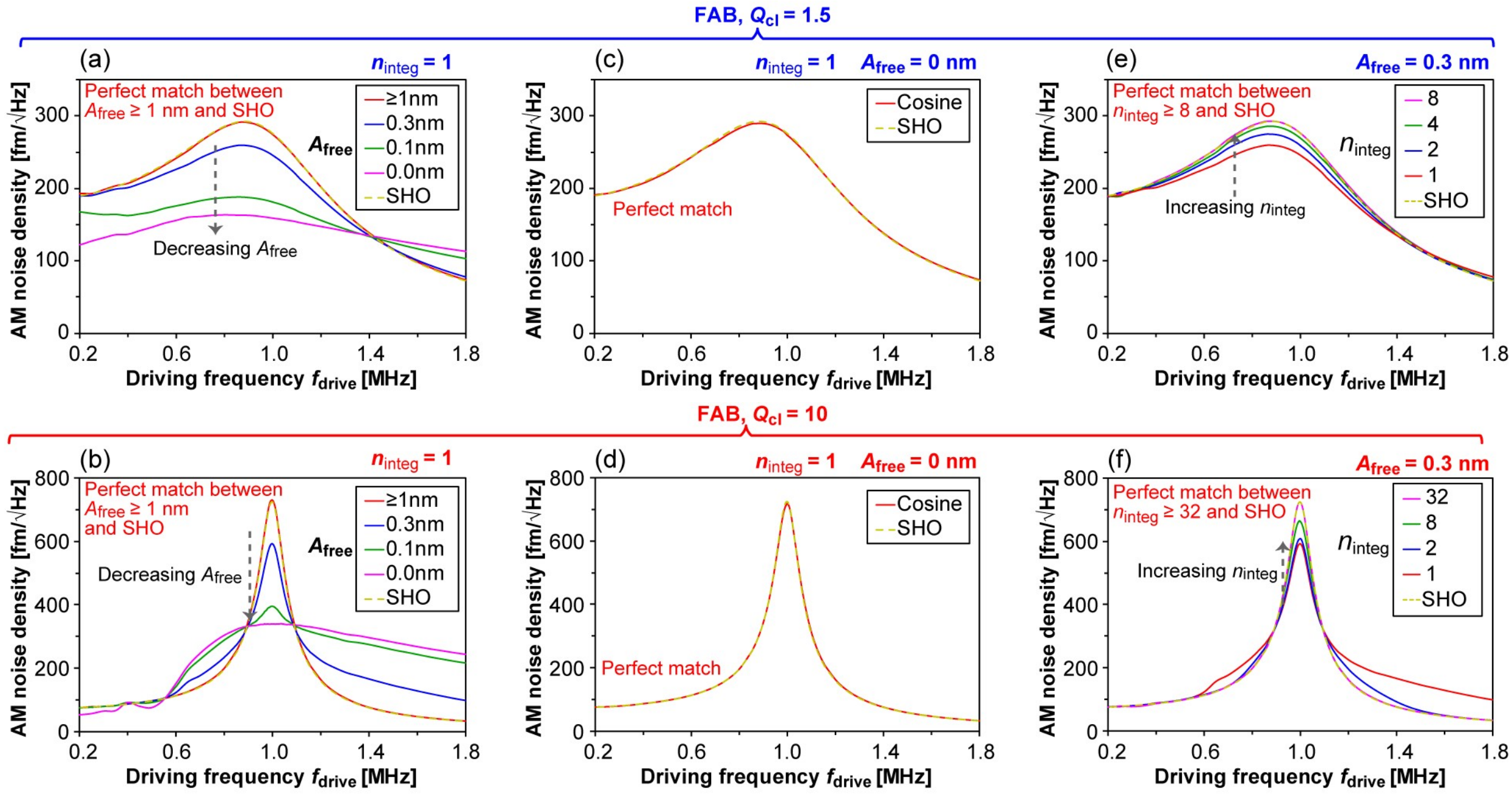

**FIG. 3.** (**a–f**) Simulated low-frequency AM noise density as a function of $f_{drive}$ for $Q_{cl}$ =1.5 (a,c,e) and $Q_{cl}$ = 10 (b,d,f). The plots include the case of $n_{integ}$ = 1 with various $A_{free}$ (a,b), the case of cosine output (c,d), and the case of $A_{free}$ = 0.3 $nm_{p-0}$ with various $n_{integ}$ (e,f).

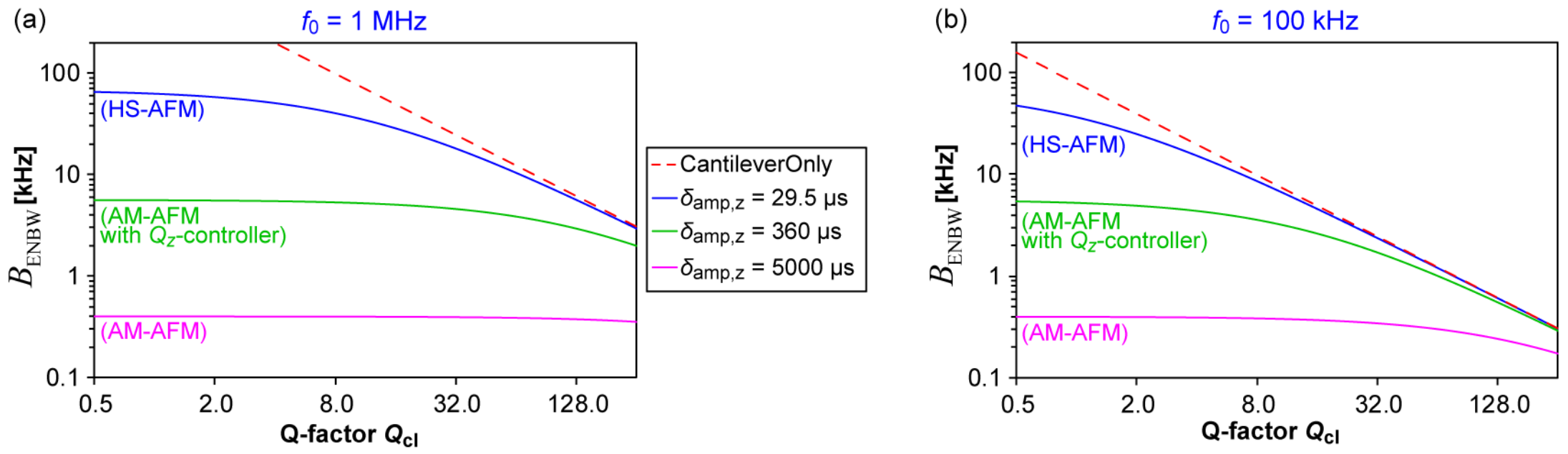

**FIG. 4.** (**a,b**) Theoretical $Q_{cl}$-dependence of $B_{ENBW}$ for cantilevers with $f_0$ = 1 MHz (a) and $f_0$ = 100 kHz (b). The blue, green, and purple curves correspond to $\delta_{amp,z}$ = 29.5 µs (typical HS-AFM), $\delta_{amp,z}$ = 360 µs (typical AM-AFM with a $Q_z$ controller), and $\delta_{amp,z}$ = 5000 µs (typical AM-AFM), respectively. The red dashed line represents the cantilever-only case.

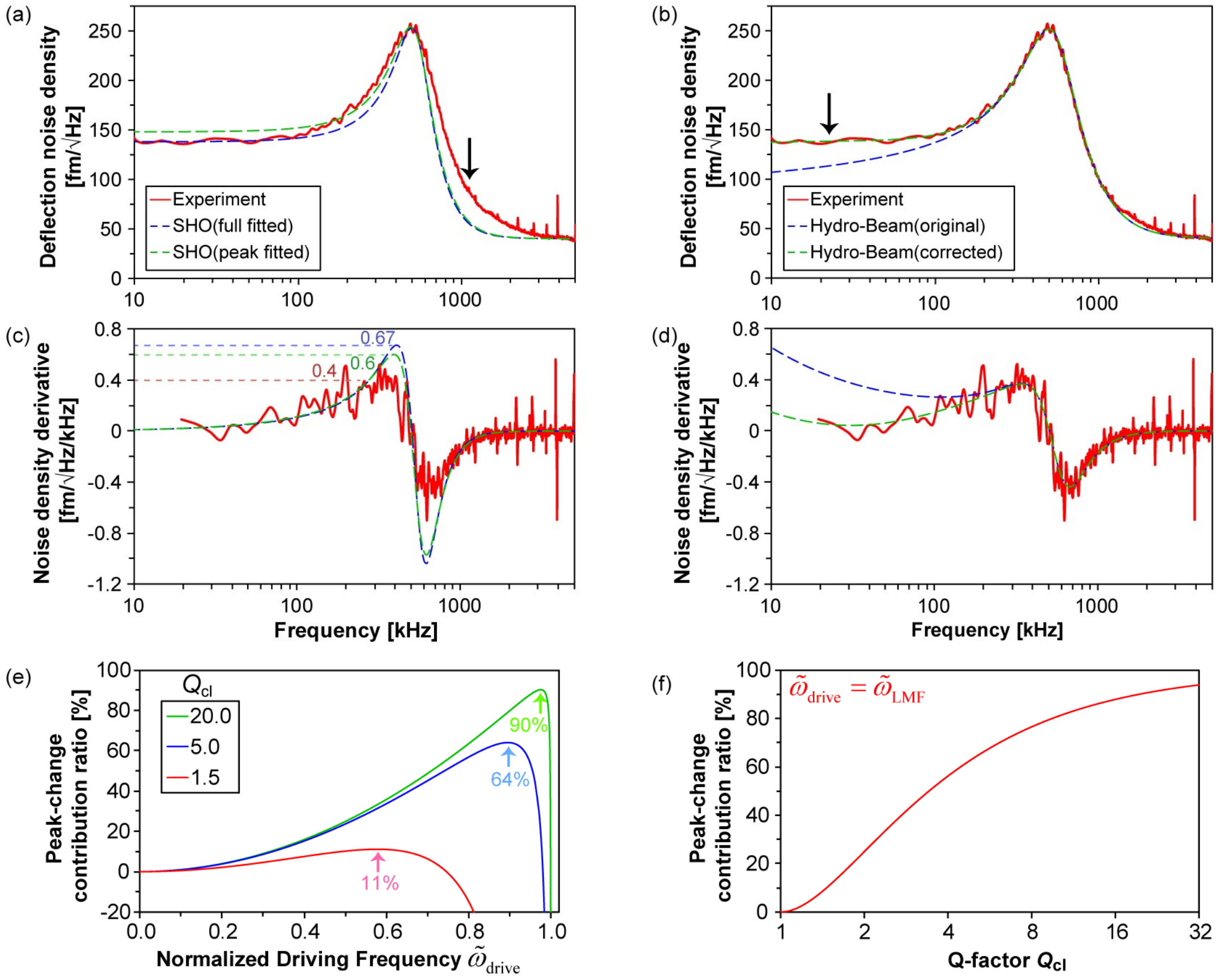

**FIG. 5.** (**a,b**) Comparison between the experimentally measured Brownian noise spectrum of the deflection and the theoretical fitting curves for the SHO (a) and Hydro-Beam (b) models. The arrows indicate regions where the discrepancy between the experimental and theoretical results is large. (**c,d**) Frequency derivatives of the experimental and theoretical curves shown in (a,b). (**e**) $f_{drive}$ dependence of the peak-change contribution to $\Delta A_{ts}$ for different $Q_{cl}$. (**f**) $Q_{cl}$ dependence of the peak-change contribution at $f_{LMF}$.

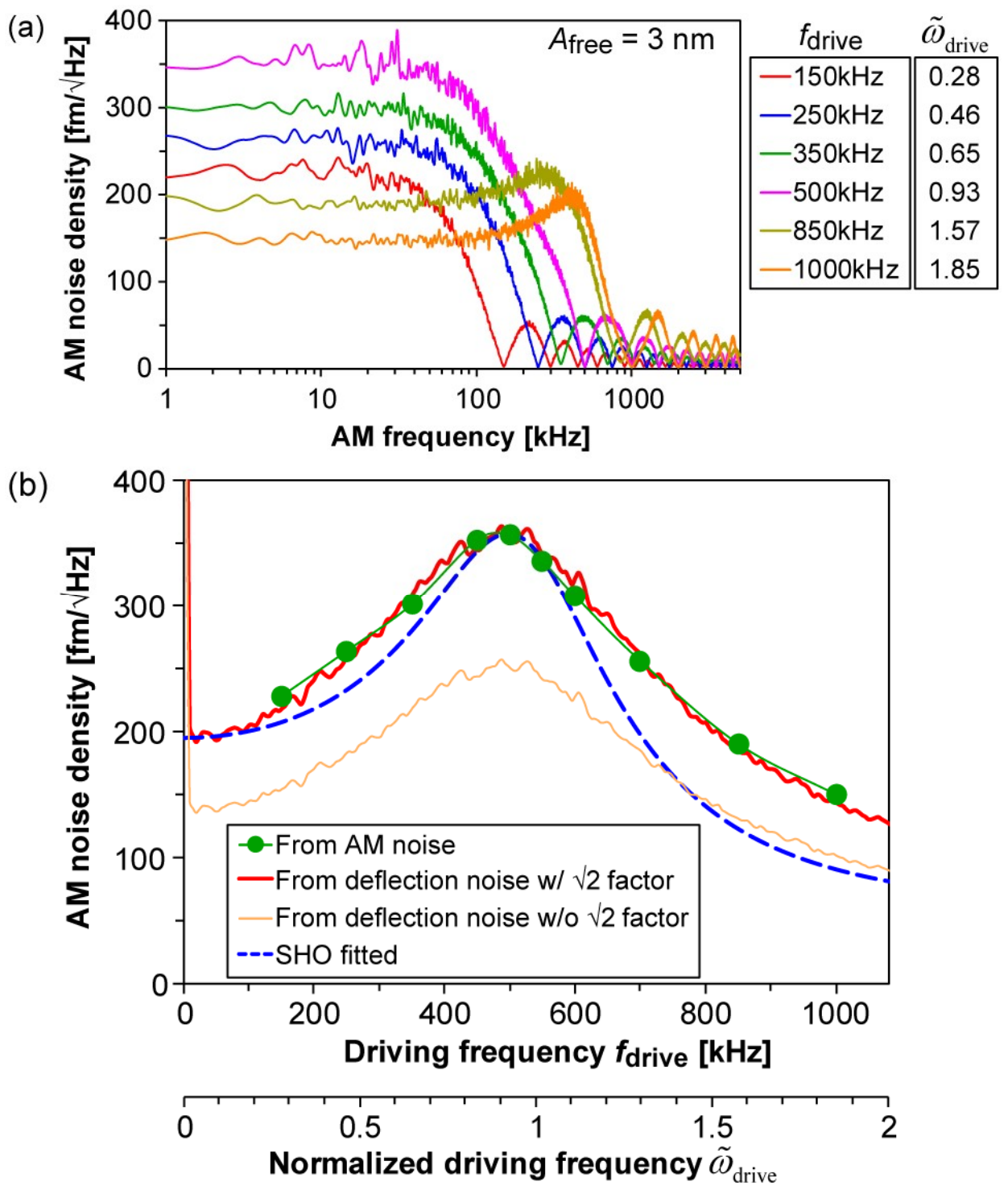

**FIG. 6.** (**a**) Experimentally acquired AM noise spectra at various $f_{drive}$ for $A_{free}$ of 3 $nm_{p\text{–}0}$. (**b**) Experimental $f_{drive}$ dependence of the AM noise density obtained from AM noise spectra shown in panel (a) (green line), from deflection noise spectra shown in Fig. 5(a) considering the $\sqrt{2}$ factor (red curve) and ignoring the $\sqrt{2}$ factor (orange curve). SHO-fitted AM noise spectra is also shown as a blue dashed curve.

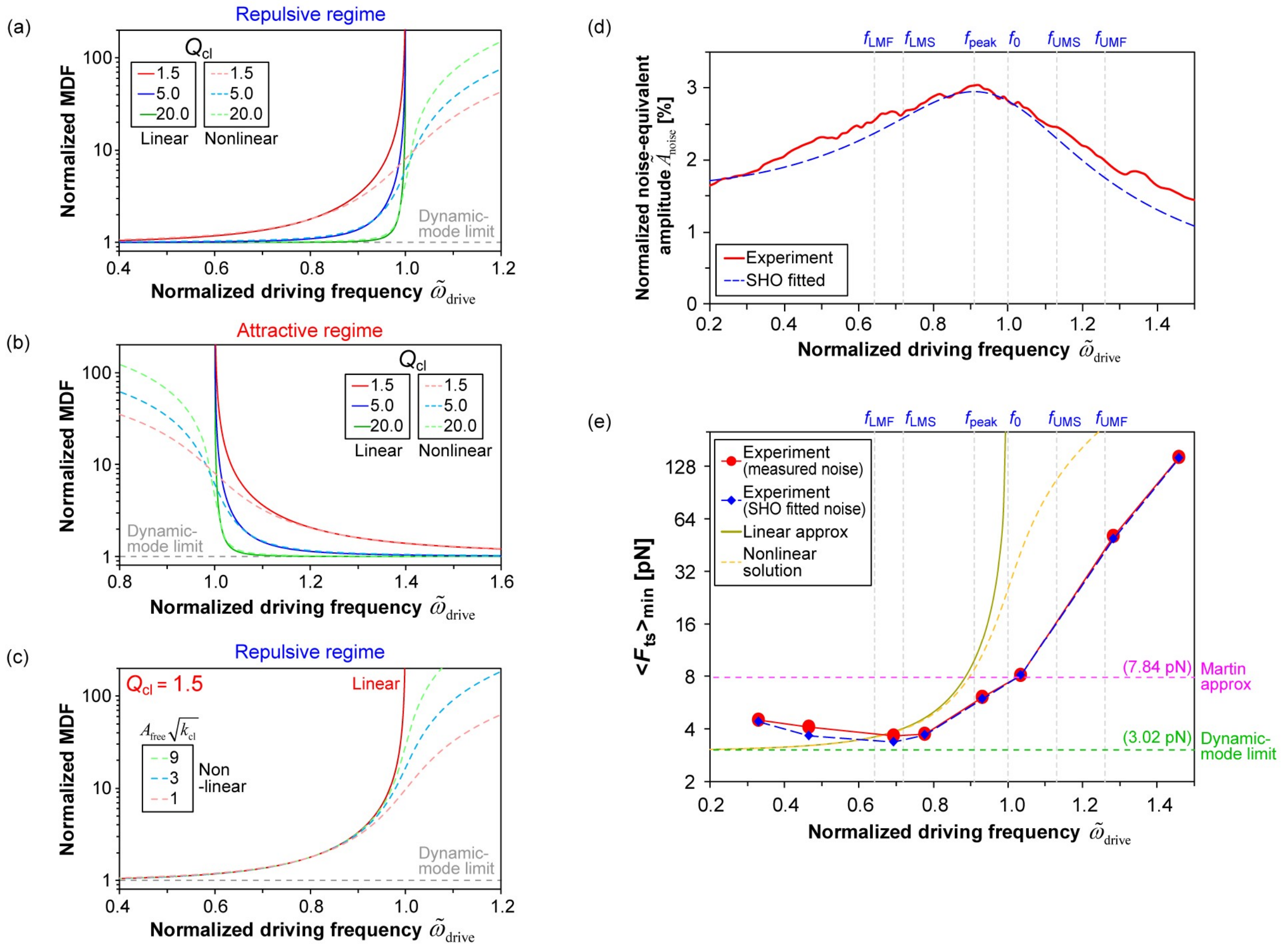

**FIG. 7.** (**a,b**) Theoretical $f_{drive}$ dependence of MDF, normalized by the dynamic-mode limit, in the repulsive (a) and attractive (b) regimes at various $Q_{cl}$. The nonlinear solutions are calculated at $A_{free}\sqrt{k_{cl}} = 1\ \mathrm{nm}\sqrt{\mathrm{N/m}}$. The solid and dashed curves represent the linear approximations and the nonlinear solutions, respectively. (**c**) Same as panel (a), but for $Q_{cl}$ = 1.5 and $A_{free}\sqrt{k_{cl}}$ values of 1, 3, and 9 $\mathrm{nm}\sqrt{\mathrm{N/m}}$. In panels (a–c), the arrows indicate the lower and upper MinForce frequencies. (**d**) Experimental $f_{drive}$ dependence of the noise equivalent amplitude normalized by $A_{free}$, expressed in percentage, along with the SHO fit result. (**e**) Comparison of the $f_{drive}$ dependence of $\langle F_{ts} \rangle_{min}^{AM}$ from experiments and analytical calculations. The force values were estimated by substituting the amplitude values obtained in panel (d) into the force–amplitude correlation measured in our previous study [59]. In panels (d,e), the vertical dashed lines indicate the characteristics frequencies.

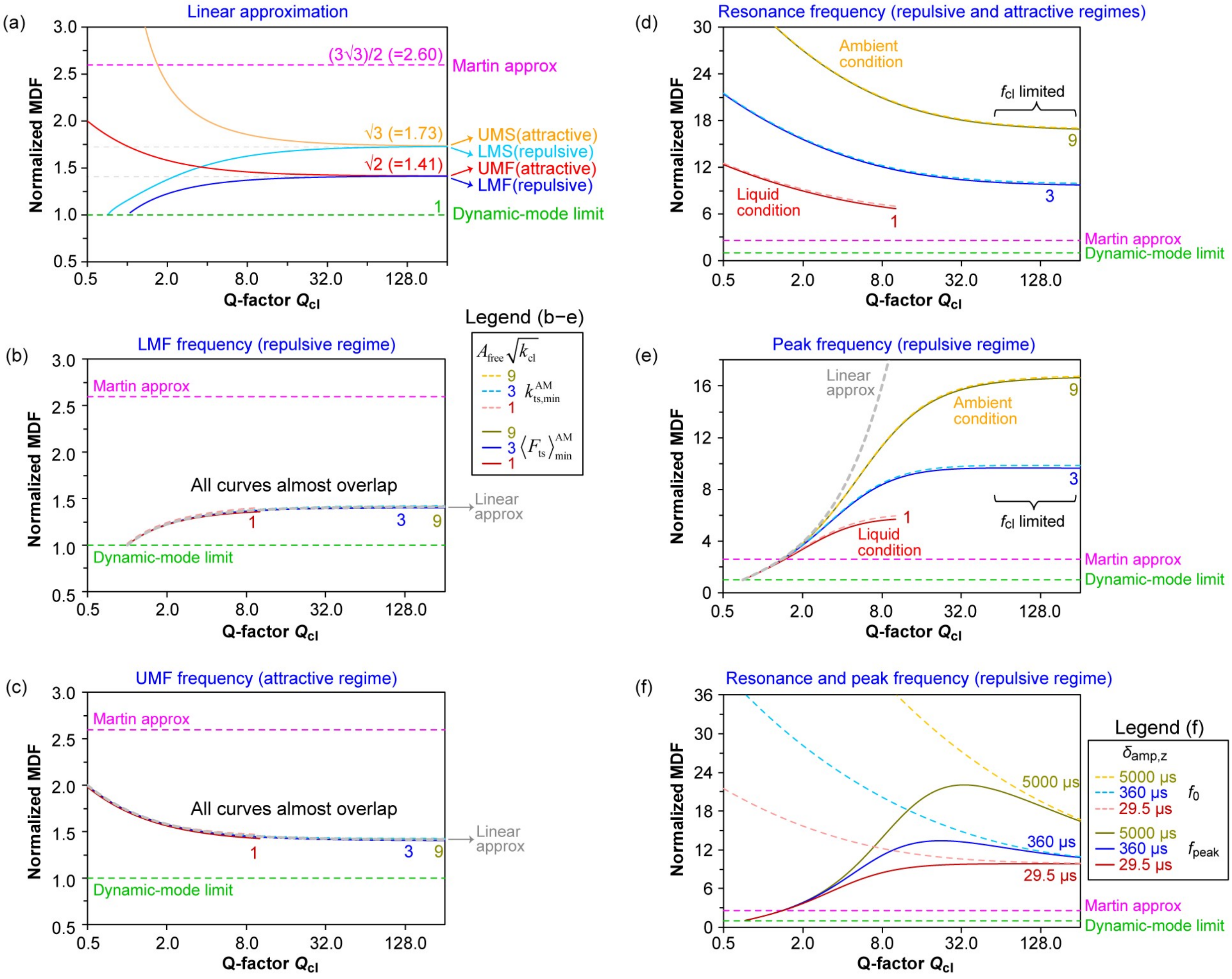

**FIG. 8.** (**a**) Theoretical $Q_{cl}$-dependence of MDF, normalized by the dynamic-mode limit, shown for the resonance slope frequencies under the linear approximation. The values shown in the graph correspond to those obtained in the limit of $Q_{cl}\to\infty$. (**b–e**) Theoretical $Q_{cl}$-dependence of the nonlinear MDF solutions, normalized by the dynamic-mode limit, for $\delta_{amp,z}$ = 29.5μs and $A_{free}\sqrt{k_{cl}}$ = 1, 3, and 9 $nm\sqrt{N/m}$, under excitation at the LMF (b), UMF (c), resonance (d), and peak (e) frequencies. The dashed and solid curves for each $A_{free}\sqrt{k_{cl}}$ represent $k_{ts,min}^{AM}$ and $\langle F_{ts}\rangle_{min}^{AM}$, respectively. The gray curves represent the linear approximation. (**f**) Theoretical $Q_{cl}$-dependence of the nonlinear MDF solutions, normalized by the dynamic-mode limit, for $\delta_{amp,z}$ = 29.5, 360, and 5000 μs under excitation at the resonance, and peak frequencies.

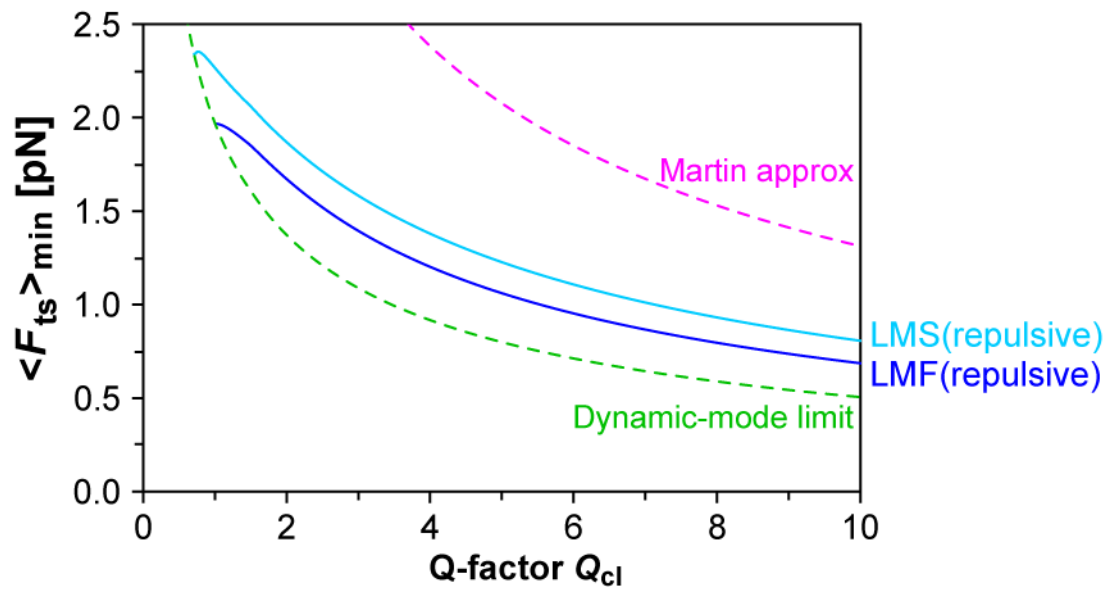

**FIG. 9.** Theoretical $Q_{\mathrm{cl}}$-dependence of explicit values of $\langle F_{\mathrm{ts}} \rangle_{\mathrm{min}}^{\mathrm{AM}}$ at the resonance slope frequencies in the repulsive regime.

## Acknowledgments

This work was supported by PRESTO, Japan Science and Technology Agency (JST) [JPMJPR20E3 and JPMJPR23J2 to K.U.]; and KAKENHI, Japan Society for the Promotion of Science [25K09575 (to K.U.), and 24H00402 (to N.K.)].

## Author contributions

K. U. constructed the theories, derived equations, and wrote the manuscript; and N. K. supervised the study.

## Data availability

The data that support the findings of this study are available from the corresponding author upon reasonable request.